\documentclass[sn-mathphys-num]{sn-jnl}
\usepackage{graphicx}%
\usepackage{multirow}%
\usepackage{amsmath,amssymb,amsfonts}%
\usepackage{amsthm}%
\usepackage{mathrsfs}%
\usepackage[title]{appendix}%
\usepackage{xcolor}%
\usepackage{textcomp}%
\usepackage{manyfoot}%
\usepackage{siunitx}
\usepackage[utf8]{inputenc}
\usepackage{newunicodechar}
\newunicodechar{₃}{$_3$}
\newunicodechar{₂}{\ensuremath{{}_2}} 

\usepackage{natbib} 
\usepackage[utf8]{inputenc}
\DeclareUnicodeCharacter{2097}{\textsubscript{l}}  

\usepackage{longtable}
\usepackage{algorithm}%
\usepackage{algorithmicx}%
\usepackage{algpseudocode}%
\usepackage{listings}%
\theoremstyle{thmstyleone}%

\theoremstyle{thmstyletwo}%
\theoremstyle{thmstylethree}%
\usepackage{amsmath,amssymb,amsfonts}
\usepackage{graphicx}
\usepackage{diagbox}
\usepackage{multirow}
\usepackage{array}
\usepackage{hhline}
\usepackage[skip=0.5\baselineskip]{caption}

\newcolumntype{P}[1]{>{\raggedright\arraybackslash}m{#1}}

\begin{document}

\title[Data-driven active learning approaches for accelerating materials discovery]{Data-driven active learning approaches for accelerating materials discovery}


\author[1,2]{\fnm{Jiaxin} \sur{Chen}}\email{chenjiaxin@nudt.edu.cn}
\equalcont{These authors contributed equally to this work.}

\author[1,2]{\fnm{Tianjiao} \sur{Wan}}\email{wantianjiaocom@nudt.edu.cn}
\equalcont{These authors contributed equally to this work.}

\author[1,2]{\fnm{Hui} \sur{Geng}}\email{genghui23@nudt.edu.cn}
\equalcont{These authors contributed equally to this work.}

\author[1,2]{\fnm{Liang} \sur{Xiong}}

\author[1,2]{\fnm{Guohong} \sur{Wang}}

\author[1,2]{\fnm{Yihan} \sur{Zhao}}

\author[1,2]{\fnm{Longxiang} \sur{Deng}}

\author[1,2]{\fnm{Zijian} \sur{Gao}}

\author[3]{\fnm{Susu} \sur{Fang}}

\author[3]{\fnm{Zheng} \sur{Luo}}

\author[1,2]{\fnm{Huaimin} \sur{Wang}}
\author*[3]{\fnm{Shanshan} \sur{Wang}}\email{wangshanshan08@nudt.edu.cn}
\author*[1,2]{\fnm{Kele} \sur{Xu}}\email{kele.xu@ieee.org}

\affil*[1]{\orgdiv{College of Computer Science and Technology}, \orgname{National University of Defense Technology}, \orgaddress{\city{Changsha}, \postcode{410000}, \state{Hunan}, \country{China}}}

\affil[2]{\orgdiv{National Key Laboratory of Parallel and Distributed Computing}, \orgname{National University of Defense Technology}, \orgaddress{\city{Changsha}, \postcode{410000}, \state{Hunan}, \country{China}}}

\affil[3]{\orgdiv{College of Aerospace Science and Engineering}, \orgname{National University of Defense Technology}, \orgaddress{\city{Changsha}, \postcode{410000}, \state{Hunan}, \country{China}}}



\abstract{
Materials discovery is a cornerstone of modern technological advancement, yet it remains constrained by traditional trial-and-error paradigms and the inherent bias of human intuition. Artificial intelligence (AI) has emerged as a transformative tool in materials science by effectively modeling structure-property relationships. Despite substantial efforts to enhance model expressiveness, data efficiency remains an equally critical challenge, given the limited availability of experimental and computational resources. Active learning (AL), as a data-driven machine learning paradigm, has shown great promise for discovering novel materials and enabling the efficient navigation of vast materials spaces. In this review, we follow the evolution of sampling strategy design techniques in AL, from Bayesian optimization to advanced deep learning-based strategies. We then highlight how AL enhances data efficiency across various data regimes, ranging from task-specific settings with limited data to the development of general-purpose datasets and large-scale models. We further provide a systematic overview of AL applications throughout the materials research pipeline, including computational simulation, composition and structural design, process optimization, and self-driving laboratory systems. Finally, we pinpoint key challenges and future perspectives of AL in materials discovery.
}


\maketitle

\section{Introduction}
\label{introduction}
Materials discovery plays an indispensable role in modern technology, underpinning advances across energy, electronics, healthcare, and information technologies. However, this process traditionally relies on extensive, time-consuming, and resource-intensive experimentation guided by human intuition, which severely limits the efficiency of new material development. Although computational techniques \cite{hachmann2011harvard, olivares2011accelerated} and high-throughput screening \cite{antoniuk2025activelearningenablesextrapolation} have enabled accelerated exploration of broader chemical spaces, such brute-force strategies remain fundamentally constrained by the substantial computational and experimental costs. Moreover, Artificial Intelligence (AI) has emerged as a transformative tool in materials science by offering an efficient approach to learning quantitative structure-property relationships from both structured and unstructured data~\cite{zhang2025artificial}. These capabilities have enabled the automation and streamlining of molecular and materials discovery across a wide range of material systems, including alloys~\cite{5,8,29}, perovskites~\cite{9,54}, and polymers~\cite{108,77}, thereby alleviating the limitations discussed above.

However, the development of robust AI solutions depends not only on sophisticated surrogate model architectures but also on the efficiency of data acquisition and iterative refinement. While substantial efforts in AI research have focused on enhancing model expressiveness and task-specific performance \cite{vaswani2017attention, devlin2019bert}, data efficiency remains an equally critical challenge, particularly in materials research. The high cost of experimental and computational validation severely limits the availability of data, leading to sparse coverage of the immense chemical and structural design space. Such data scarcity not only limits model generalizability but also biases learning toward well-explored regions. This raises a central question: \textbf{how can we improve data efficiency in AI-driven materials research while respecting practical constraints on experimental and computational resources?}
\begin{figure}[h]
\centering
\includegraphics[width=1\linewidth]{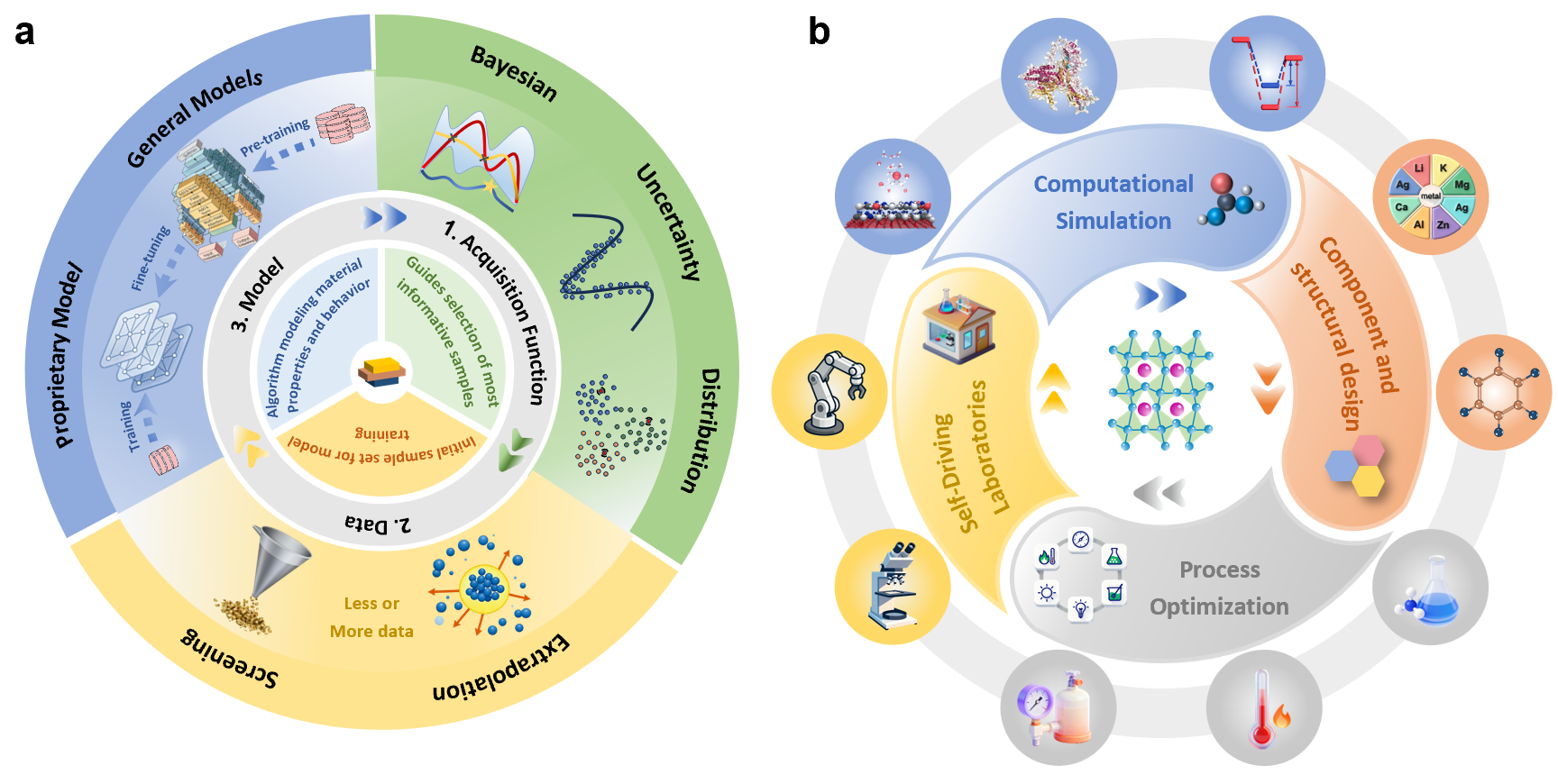}
\caption{\textbf{Schematic of the active learning framework for accelerating materials discovery.} \textbf{a} The active learning (AL) framework integrates three essential components: models for prediction, data for training, and acquisition functions for sample selection. \textbf{b} The application landscape of AL across the materials research workflow. The cycle illustrates four key stages: accelerating computational simulation, guiding compositional and structural design, optimizing parameters through process optimization, and enabling autonomous execution within self-driving laboratories.} \label{fig:fig1}
\end{figure}

Active learning (AL) addresses this challenge through a data-driven paradigm that iteratively identifies informative data points to investigate in the candidate material space~\cite{205,wan2023survey, tharwat2023survey}. By shifting from a human-centered trial-and-error paradigm to a data-driven, AI-centric paradigm, it holds great promise for accelerating materials screening and even for guiding discovery beyond existing domain knowledge. Specifically, AL employs a sampling strategy to select the next material to be evaluated by an oracle. Then, the newly acquired data are incorporated to update the model and inform subsequent sampling in an iterative, closed-loop manner. Instead of exhaustively evaluating all possibilities in the entire chemical space, AL substantially improves data efficiency and enables efficient exploration of vast chemical spaces.

Over the past decade, AI powered by AL methodologies has emerged as a transformative paradigm in materials discovery. Classical AL approaches, such as Bayesian optimization~\cite{3,4,8}, have been widely applied in data-scarce regimes to guide experimental efforts toward regions of materials space that are expected to exhibit higher target properties. Moreover, AL continues to evolve alongside advances in deep learning (DL) architectures and large-scale foundation models. By incorporating uncertainty estimation, feature distribution-aware sampling, and task-dependent objectives, AL has been explored as a framework for guiding decision-making in various research workflows~\cite{127,128,134,135,137}. More recently, AL has been investigated as a tool for constructing datasets that support the pre-training or fine-tuning of general-purpose materials models~\cite{yin2025surff,22,11}. Together, these developments suggest a potential role for AL in bridging task-specific studies and broader modeling efforts, although challenges related to robustness, scalability, and evaluation remain.

In this review, we provide a critical overview of AL in data-driven materials discovery. We systematically review various AL methods, highlighting their integration with traditional machine learning (ML) approaches and modern deep learning frameworks (Fig.~\ref{fig:fig1}a). Then, we examine how AL has been used to improve data efficiency across different model and data regimes, encompassing task-specific models trained on limited data and larger models designed for more generalizable datasets. We further summarize representative applications of AL throughout the materials research pipeline, including computational simulation, compositional and structural design, process optimization, and self-driving laboratory systems (Fig.~\ref{fig:fig1}b). Finally, we discuss the challenges specific to AL in materials research, such as the cold-start problem, the integration with prior knowledge, and the choice of appropriate AL configurations, and outline directions for future work aimed at improving the robustness and accessibility of AL tools.

\section{Active learning approach}\label{method}

\begin{figure}[h]
\centering
\includegraphics[width=1\linewidth]{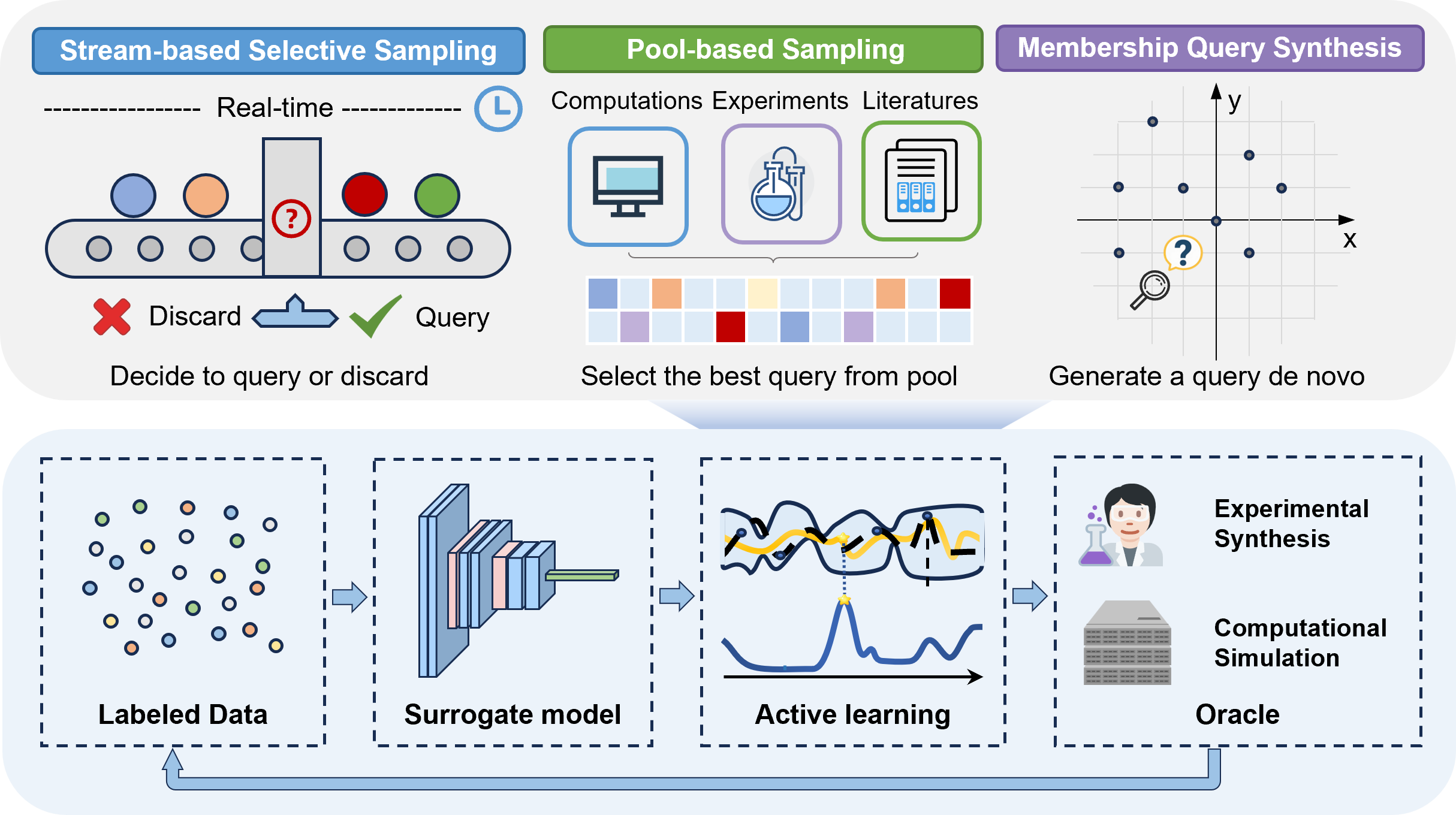}
\caption{\textbf{The closed-loop workflow of active learning.} The process begins with a candidate data pool and labeled data, which is used to train the surrogate model. Through an AL acquisition strategy, the model identifies the most promising candidates for evaluation. These candidates are subsequently evaluated by an oracle (e.g., experimental synthesis and characterization, or computational simulation), with the newly acquired data fed back to iteratively refine the model, creating a closed-loop loop that efficiently accelerates discovery. Three representative sampling scenarios are illustrated: stream-based selective sampling, where the model makes real-time decisions to query or discard incoming data points; pool-based sampling, involving selection of the most promising queries from a predefined candidate pool; and membership query synthesis, where the learner generates new queries at specific coordinates in the input space to maximize information gain.}
\label{fig:fig4}
\end{figure}

Classic AL scenarios are primarily categorized into three scenarios: stream-based selective sampling, pool-based sampling, and membership query synthesis (Fig.~\ref{fig:fig4})~\cite{205}. In materials science, these paradigms manifest with distinct impact. The classic scenario for materials discovery remains firmly grounded in pool‑based sampling, a strategy that iteratively selects the most informative candidates from a predefined data pool, such as a combinatorial library of compounds. Stream‑based selective sampling, although powerful for real‑time decision‑making, sees limited application in materials research, primarily in computational settings such as on‑the‑fly learning during molecular dynamics simulations. In contrast, membership query synthesis has emerged as a emerging frontier. It forms the algorithmic core of self-driving laboratories, allowing the learner to request labels for any instance in the input space, including queries generated de novo rather than drawn from an existing distribution. This capability enables the autonomous design of unprecedented experimental parameters and material compositions, thereby moving beyond the constraints of static candidate pools.

\begin{figure}[h]
\centering
\includegraphics[width=1\linewidth]{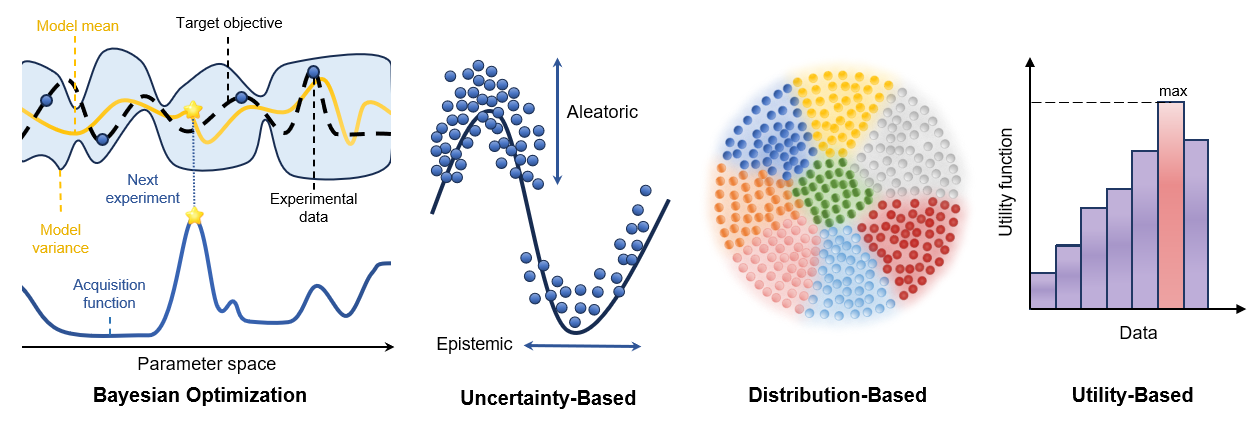}
\caption{\textbf{Diverse active learning approaches.}  Four representative AL strategies: Bayesian optimization navigates the parameter space through acquisition functions to balance exploration and exploitation; uncertainty-based approaches select samples based on various uncertainty metrics; distribution-based methods capture the overall data distribution through representative sampling; and utility-based strategies prioritize samples by their expected utility.} \label{fig:fig3}
\end{figure}

To implement this workflow, AL approaches are generally divided into traditional AL and deep active learning (DAL). Traditional AL methods have demonstrated strong effectiveness in settings with limited training data and fixed feature representations. With the advent of deep neural networks (DNNs) and their powerful representation-learning capabilities, DAL has emerged as a natural extension of AL. In practice, DAL not only adapts classical AL heuristics to deep models but also introduces novel strategies tailored specifically to DNN architectures. Consequently, the development of effective AL strategies in materials science necessitates the integration of the underlying methodology with domain-specific knowledge. Figure~\ref{fig:fig3} illustrates the key categories of AL approaches in materials research, spanning from traditional to deep learning methodologies. The following subsections provide a detailed overview of these approaches, outlining both their algorithmic foundations and their applicability to materials-specific challenges.

\subsection{Traditional active learning approach}
Traditional AL methods typically utilize classical ML models such as Gaussian processes (GPs), random forests (RFs), or support vector machines (SVMs). These approaches rely on manually engineered feature representations, such as Smooth Overlap of Atomic Positions (SOAP)~\cite{231,14,18}, Simplified Molecular Input Line Entry System~\cite{weininger1988smiles}, Extended-Connectivity Fingerprints~\cite{rogers2010extended}, Atom-Centered Symmetry Functions (ACSF)~\cite{behler2011atom}, and descriptors derived from domain-specific knowledge~\cite{29,54}. By utilizing these fixed, pre-processed features, traditional AL methods aim to efficiently explore the data space by selecting the most representative or informative samples for labeling, thereby maximizing performance with limited resources. In data-scarce regimes, traditional methods often remain the preferred choice, offering superior stability and interpretability due to their reduced complexity and effective integration of domain-specific features. Traditional AL methods are generally categorized into Bayesian optimization (BO), uncertainty-based methods, distribution-based methods, utility-based methods, hybrid methods, and others.

\subsubsection{Bayesian optimization}
The structure-property relationships in materials science are often poorly understood, and their modeling and optimization are typically treated as black-box problems. BO provides a robust framework to address such challenges (Fig.~\ref{fig:fig3}a). BO aims to balance the exploration of uncertain regions with the exploitation of promising regions, thereby efficiently navigating the vast materials space with limited experimental resources to identify promising candidates. It is particularly effective for optimizing expensive black-box functions and has been shown to surpass human baselines in specific materials applications~\cite{217,218}. Consequently, owing to its capacity to handle broad exploration spaces encompassing varied chemical parameters and its support for multiple parallel experiments, BO has been extensively adopted in the field of materials informatics~\cite{94,108,109,124,134,171,219,179,181,220,221,222}.

For a pre-defined search space, BO iteratively trains a probabilistic surrogate model to approximate the target function, with an acquisition function guiding the search by balancing exploitation and exploration. Exploitation refines existing knowledge by focusing on high-performing regions, while exploration expands understanding by sampling uncertain or diverse areas. Striking the right balance between these strategies is essential for efficiently directing experimental workflows toward promising materials. Common acquisition functions include Thompson Sampling (TS), Probability of Improvement (PI), Expected Improvement (EI), and Upper/Lower Confidence Bound (UCB/LCB).

\textbf{Thompson sampling.} TS is a randomized acquisition strategy that balances exploration and exploitation by sampling from the posterior distribution of predictions~\cite{225}. In each iteration, a stochastic realization $f_s(x)$ is drawn from the GP posterior. The point that maximizes this sampled function is selected as the recommended candidate~\cite{171}. The acquisition function is defined as
\begin{equation}
    TS(x) = f_s(x),
\end{equation}
where $x$ is the candidate material, and $f_s(x) \sim \mathcal{N}(\mu(x), \sigma^2(x))$, with $\mu(x)$ and $\sigma^2(x)$ representing the predicted mean and variance for $x$ by the GP model.

\textbf{Probability of improvement.}
Improvement-based acquisition functions tend to prioritize candidate materials expected to surpass the current best-performing candidate~\cite{217}. Specifically, PI evaluates the probability that a candidate material $x$ will outperform the current best, $f(x^+)$. Formally, it is defined as:
\begin{equation}
    PI(x) = \Phi(\frac{\mu(x) - f(x^+) - \epsilon}{\sigma(x)}),
\end{equation}
where $f(x^+)$ is the incumbent maximum observed performance, $\sigma(x)$ is the predicted standard deviation, and $\Phi( \cdot )$ represents the standard normal cumulative distribution function. The parameter $\epsilon$ controls the exploitation-exploration balance: smaller $\epsilon$ favors local refinement of known regions, while larger $\epsilon$ promotes exploration of uncertain regions in the material space.

\textbf{Expected improvement.}
While PI is effective for target-oriented searches, it often leads to myopic exploitation by focusing solely on the probability of improvement while neglecting the magnitude of potential gain, causing the algorithm to stagnate in local minima. To address this limitation, EI~\cite{226} considers both the probability and magnitude of performance gain, denoted as $I(x)$:
\begin{equation}
    I(x) = (f(x) - f(x^+))\mathbb{I}(f(x) > f(x^+)),
\end{equation}
where $\mathbb{I}(\cdot)$ denotes the indicator function.
Given the normality of $f(x)$ under the GP predictive distribution, the expectation can be computed analytically as follows~\cite{227,217,219}:
\begin{equation}
EI(x) = \begin{cases}
(\mu(x) - f(x^+) - \epsilon) \Phi(Z) + \sigma(x) \phi(Z), & \text{if } \sigma(x) > 0, \\
0, & \text{if } \sigma(x) = 0.
\end{cases}
\end{equation}
where $Z = \frac{\mu(x) - f(x^+) - \epsilon}{\sigma(x)}$, and $\phi(\cdot)$ represents the probability density function of the standard normal distribution.
These improvement strategies have been extensively validated across numerous previous studies, demonstrating their ability to efficiently guide the discovery of new materials~\cite{92,147,102,15,22,35,114,134,174,81,173,94,3,175,26,89,176,67,47}.

\textbf{Upper and lower confidence bounds.}
Unlike improvement-based strategies, UCB and LCB employ a confidence interval approach, guiding the search by considering the upper and lower bounds of the expected material performance. These approaches offer a more intuitive trade-off mechanism as follows~\cite{230}:
\begin{equation}
UCB(x) = \mu(x) + \kappa \sigma(x),
\end{equation}
\begin{equation}
LCB(x) = \mu(x) - \kappa \sigma(x),
\end{equation}
where $\kappa$ is a hyperparameter that controls the exploration-exploitation balance and the linear combination explicitly incorporates the predicted mean and standard deviation~\cite{26,49,81,91,108,137,147,170,173}.

\textbf{Multi-objective Bayesian optimization.} Traditional BO is primarily designed for single-objective optimization, limiting its capacity to navigate the multi-objective trade-offs inherent in materials science, such as the classic conflict between strength and ductility~\cite{12}, electrical conductivity and thermal stability~\cite{179}, or the balance between reaction yield and selectivity~\cite{3}. In multi-objective Bayesian optimization (MOBO), the objective shifts from locating a single optimum to identifying a Pareto-optimal set representing the most favorable trade-offs among competing metrics. Therefore, traditional acquisition strategies, designed for single-objective settings, are not directly applicable in multi-objective scenarios and require specific adaptations or extensions. Early MOBO approaches primarily adapted the EI framework. One strategy employs scalarization by constructing unified utility functions~\cite{89}, weighted EI summation, or constrained improvement methods~\cite{99}. Another strategy computes EI for each objective independently, followed by Pareto selection~\cite{176}.
Unlike these EI-based methods, Expected Hypervolume Improvement (EHVI) offers a parameter-free alternative that directly measures the expected increase to the hypervolume dominated by the Pareto front, eliminating the need for pre-defined weights~\cite{3,12,171}. Its batch extension, qEHVI, enables parallel evaluation of multiple candidates for high-throughput materials discovery~\cite{179,175,124}.

\subsubsection{Uncertaint-based approach}
The materials discovery process typically focuses on identifying materials with superior properties, often extending beyond the boundaries of existing domain knowledge. In this context, uncertainty-driven AL methods play a crucial role by systematically prioritizing samples that exhibit high uncertainty based on the surrogate model's information~\cite{203} (Fig. \ref{fig:fig3}b). These strategies not only effectively reduce predictive uncertainty in models but also facilitate efficient exploration within the vast material space, thereby accelerating the materials discovery process~\cite{204}.

\textbf{Entropy.}
Entropy, a fundamental concept in information theory~\cite{233}, is commonly used to quantify the uncertainty of a predictive distribution. It is defined as:
\begin{equation}
    H(x)=  -\sum_{i} P(y_i|x; \theta) \log P(y_i|x; \theta),
\end{equation}
where $y_i$ encompasses all possible class labels and $\theta$ represents the model parameters. By maximizing Shannon entropy, algorithms can identify the most informative samples. This approach is widely adopted in frameworks such as BO, thereby guiding model training toward regions of high informational gain and enhancing predictive performance~\cite{34,128,205}.

\textbf{Least confidence.}
While entropy accounts for the full probability distribution, the least confidence (LC) criterion serves as a simpler alternative by focusing exclusively on the most probable prediction. LC selects samples with the lowest posterior probability of the predicted label $\hat{y}$~\cite{95,206}:
\begin{equation}
    LC(x) = 1 - P(\hat{y} | x; \theta),
\end{equation}
where $\hat{y} =\arg\max_{y}P(y|x;\theta)$ is the predicted class for input $x$.

\textbf{Predictive variance.}
Predictive Variance (PV) is also a widely adopted uncertainty-based sampling strategy in AL, frequently employed in the exploration stage of BO~\cite{4,25,30,33,38,84,93,96,110}. PV selects samples that maximize the PV of the model, based on the premise that higher variance reflects greater predictive uncertainty:

\begin{equation}
PV(x)=\sigma^2(x).
\end{equation}

\textbf{Query by committee.}
Beyond methods based on a single probabilistic model, uncertainty can also be assessed using the Query by Committee (QBC). This approach constructs a committee composed of multiple learners, referred to as members, which collectively predict the labels of candidate samples, rather than relying on a single model. Within this framework, committee members provide predictions for each query sample, and the most informative samples are identified as those eliciting the greatest disagreement among members~\cite{111,28}. Notably, ensemble models such as RFs are inherently compatible with QBC mechanism and can be readily integrated into this framework~\cite{50,64}.

\textbf{Other uncertainty strategies.} Furthermore, researchers have expanded the AL sampling framework through diverse perspectives, developing more varied strategies to improve query efficiency and model generalization capabilities. For instance, a Gaussian-weighted heuristic has been developed to prioritize candidates stochastically around a target value, moving away from standard variance-based methods. This approach can be validated using a rolling forecasting origin framework, which evaluates performance by iteratively retraining the model on cumulative historical data to predict the immediate subsequent batch, thereby emulating the chronological progression of experimental discovery~\cite{8}. Similarly, Convex Hull-Aware AL relies on GPR to model energy surfaces and guides experimental selection by reducing uncertainty relative to the convex hull~\cite{44}. Moving beyond static uncertainty estimation, the HyperAL method intervenes in the data generation process by introducing bias terms into MD simulations, thereby actively steering the system to explore high-uncertainty configurations~\cite{83}. In addition, integrating semi-supervised learning with fundamental physical constraints, such as the Gibbs phase rule, allows for the use of label propagation to quantify uncertainty and significantly improves the efficiency of phase diagram construction~\cite{127}. Finally, for complex properties that elude purely algorithmic definition, human expertise can be integrated into the loop. Molecular complexity evaluation is optimized by selectively calibrating uncertain pairs based on expert feedback, enhancing labeling efficiency by focusing on subtle differences that models might overlook~\cite{145}.

\subsubsection{Distribution-based approach}
In contrast to methods that directly quantify model uncertainty,
distribution-based approaches focus on understanding the underlying structure of the data and selecting batches of samples that best represent the unlabeled data distribution. The key idea is that these representative examples, once labeled, can adequately represent the distributional properties of the entire dataset

Distribution-based sampling performance depends significantly on the quality of the material space representation. As a widely used descriptor, SOAP~\cite{231} converts atomic coordinates into physical descriptors that are invariant to rotation, translation, and permutation, enabling robust descriptions of the local chemical environment~\cite{14,33,18,93,96,130}. Leveraging such representations, advanced distribution-based data selection strategies have been developed. One prominent approach integrates SOAP descriptors with metrics for maximum and average interatomic distances to assess the representativeness of new structures. By setting thresholds to balance similarity and diversity, a substantial increase in structural coverage across the material space can be achieved~\cite{14}. Similarly, SOAP-based similarity selectors can be combined with distance-based selectors utilizing the local outlier factor to effectively identify diverse samples in high-dimensional feature spaces~\cite{18}. Additionally, a set of descriptors collectively termed CBAD, which includes Cell parameters, Bonds, Angles, and Dihedrals, has been proposed to quantitatively characterize local structures and screen for atomic configurations with unique properties~\cite{41}.

Building upon these representations, distribution-based methods employ tailored algorithm for the efficient exploration of high-dimensional landscapes. Clustering-based sampling partitions data into groups based on inherent structural patterns~\cite{35}. K-means divides data into K predefined clusters using distance metrics~\cite{232,35}, while density-based algorithms such as Hierarchical Density-Based Spatial Clustering of Applications with Noise (HDBSCAN) automatically determine cluster numbers and detect arbitrarily shaped clusters with varying densities. For instance, integrating HDBSCAN with MD sampling extracts representative configurations for training interatomic potentials, substantially reducing reliance on expensive AIMD calculations~\cite{130,135}. Complementing these geometry-based approaches, causal-driven AL frameworks construct causal graphs across material subspaces and identify data subsets that mirror the global causal structure, efficiently reconstructing structure-property relationships~\cite{141}.

Despite thir success, the aforementioned approaches are often hampered by their reliance on fixed representations, which limits their adaptabilityand fail to adapt to shifting data distributions during the learning process. To address this challenge, dynamic feature adjustment methods have been proposed~\cite{146,112}. One such strategy is the Feature-Adaptive BO framework~\cite{146}, which adaptively selects optimal feature combinations from a predefined global set. By actively eliminating redundancy via algorithms such as minimum Redundancy Maximum Relevance (mRMR), this approach dynamically updates the material representation and significantly accelerates the BO search. To move beyond fixed feature sets entirely, another method generates a dynamic pool of molecular fragments known as the Amon pool~\cite{112}. During optimization, the most relevant Amons are extracted for training, and the pool expands progressively. Notably, this approach achieves precision comparable to conventional methods utilizing thousands of molecules while requiring only a few dozen Amons.

\subsubsection{Utility-based approach}
Utility-based methods represent a goal-oriented sampling strategy in which the sampling process is explicitly aligned with material design objectives. Typically, this involves defining a utility function to identify candidate samples that maximize the expected value of target properties. Such approaches are particularly well suited to materials discovery and optimization tasks with clearly defined objectives and a known search space, as they effectively enhance experimental efficiency.

Several studies have designed targeted utility functions to enable goal-oriented sampling in specific scenarios. For instance, predicted values are often directly employed as the utility criterion to select samples with the highest performance or those approaching expected target values, thereby guiding the search toward optimal regions~\cite{5,169,181}.
In addition, a utility evaluation criterion centered on the probability of reaching the target value has been proposed to provide new insights for optimization~\cite{183}.
Beyond simple predictions, utility functions incorporating metrics such as liquid density error or radial distribution function deviations have been designed to construct optimization objectives with physical significance~\cite{159}.
Furthermore, the integration of a transfer learning mechanism with domain knowledge within the utility function has been utilized to enhance learning efficiency for the target task~\cite{119}.
Despite this progress, utility-based approaches, particularly those prioritizing exploitation, rely heavily on model performance. By favoring samples predicted to be optimal by the current model, these methods often concentrate sampling in known high-performance regions, increasing the risk of converging to local optima. Consequently, purely utility-driven strategies are frequently integrated into the exploitation phase of AL frameworks rather than serving as standalone global search strategies.

\subsubsection{Hybrid approach}
While uncertainty, data distribution, and utility-based AL strategies have offered valuable insights into solving material science problems, each faces inherent limitations.
For example, uncertainty sampling may select redundant samples and is sensitive to noise and outliers in experimental data, potentially over-prioritizing anomalous points.
Distribution-based methods, although capable of covering the design space, often select numerous suboptimal samples that contribute little to performance improvement, thereby consuming substantial experimental resources.
Utility-based methods often rely on task-specific functions, which can limit their transferability across different systems.
Consequently, hybrid strategies have been developed to combine the strengths of multiple approaches, mitigating issues such as bias and sample redundancy that arise from relying on a single method.

One typical hybrid strategy employs a multi-stage approach, dividing the sampling process into distinct phases that utilize different AL criteria. Clustering-based strategies are often effectively combined with measures such as uncertainty or EI to enhance diversity while boosting information gain~\cite{76,58,72}. Alternatively, multiple optimization criteria can be unified within a single acquisition function that balances them at each selection step. One method leverages the hierarchical structure of decision trees to partition the data space and dynamically balance both uncertainty and diversity across each region~\cite{56}. Evolutionary principles have also been embedded within the AL mechanism to enhance exploration. For instance, the Adaptive Learning Genetic Algorithm utilizes crossover and mutation to efficiently explore experimental conditions~\cite{166}, while the integration of evolutionary heuristic search with BO's probabilistic modeling has been shown to substantially improve the discovery of the Pareto front under constraints~\cite{10}. Through the synergistic integration of diverse criteria, these hybrid frameworks transition from narrow exploitation toward robust, multi-dimensional search, laying the foundation for navigating increasingly complex and high-dimensional material space.

\subsection{Deep active learning approach}
Benefiting from the powerful representation learning of over-parameterized architectures, DNNs have been increasingly applied to various materials science tasks. While DNNs achieve impressive performance by leveraging large-scale training datasets, the substantial cost of data acquisition remains a significant constraint. To address this, DAL integrates DL with AL to improve data efficiency and model performance under limited labeling budgets. Conceptually, DAL extends the principles of AL from traditional ML tasks, which has been extensively studied in fields such as computer vision and natural language processing in recent years~\cite{206, 234}. The taxonomy of sampling strategies in DAL is largely consistent with that in traditional ML tasks. While several tradirional AL methods have been directly adapted to DL, numerous novel approaches specifically tailored to DNNs have also been proposed.

\subsubsection{Deep Bayesian optimization}
BO, as a foundational traditional AL method, has also been extended to DL models. However, a major challenge lies in the fact that traditional BO relies on models, such as GPs, for mathematically grounded uncertainty quantification. This approach struggles with high-dimensional, discontinuous, and non-stationary data, where DL is often required. To address this and adapt BO for DL, deep kernel learning~\cite{calandra2016manifold, wilson2016deep, wilson2016stochastic} integrates neural network representation learning with GP-based uncertainty quantification. One implementation involves combining convolutional neural networks with GPR models by using the output from the penultimate layer of the pre-trained GoogLeNet as input for GPR model training. Despite its promise in materials science, deep kernel learning remains constrained by the poor scalability of GPs in high-dimensional feature spaces, risks of mode collapse, and conflicting optimization dynamics between its neural network and GP components~\cite{ober2021promises}.

Bayesian neural networks (BNNs), in which all network weights are treated as probability distributions rather than deterministic scalar values, provide a principled framework that combines powerful representation learning capabilities with reliable uncertainty quantification~\cite{titterington2004bayesian, lampinen2001bayesian}. However, reliable Bayesian inference requires computationally intensive sampling methods, making full BNNs prohibitively expensive. Monte Carlo Dropout~\cite{gal2016dropout} provides a computationally efficient and practical alternative to BNNs for uncertainty quantification. This approach has been integrated with a graph convolutional neural network with Monte Carlo Dropout as an approximate Gaussian process to identify promising candidates. Other strategies to mitigate computational costs include constructing surrogate models from parameter kernel densities estimated via BNNs~\cite{185} or employing partial BNNs~\cite{165}. By making strategic choices about which layers are treated probabilistically, this design achieves uncertainty estimates and predictive accuracy in AL tasks comparable to fully Bayesian networks, but at a substantially reduced computational overhead.

In addition, various BO strategies are employed in DL to select samples, with the EI strategy~\cite{47,92,120} being the most commonly used. For example, researchers have applied the EI strategy to select 50 polyimides with high fractional free volume that were most likely to enhance performance in each experiment for validation~\cite{92}. Other implementations include dynamic threshold Bayesian AL, where DFT calculations are triggered by Graph Neural Network (GNN) prediction errors or predefined step limits~\cite{143}.
Beyond identifying a single optimum, recent variants focus on discovering diverse and high-value candidates through frameworks such as multi-fidelity AL combined with generative flow networks~\cite{163}.
Furthermore, MOBO has also been integrated with DL to identify a set of Pareto-optimal solutions, balancing multiple desired goals. A representative implementation of this approach involves screening organic molecules for high theoretical energy density and low energy gaps~\cite{47}. This multi-objective AL framework utilizes a hybrid ML model that combines convolutional neural networks with GPR to predict properties, while the Non-dominated Sorting Genetic Algorithm II is employed to identify the Pareto front across the target property space.

\subsubsection{Deep uncertainty-based approach}
Deep uncertainty-based query strategies exploit the uncertainty estimates provided by models to prioritize samples that are least reliable, thereby guiding experiments toward unexplored regions of the material space. However, applying these strategies in DAL presents unique challenges compared to traditional AL. Traditional probabilistic models, such as GPs, offer intrinsic and mathematically grounded uncertainty estimates. In contrast, standard DNNs are inherently deterministic point estimators that frequently suffer from overconfidence, often assigning high confidence scores even to predictions far from the training distribution~\cite{wang2014new, nguyen2015deep}. Consequently, implementing uncertainty sampling in the DL regime requires not only adapting traditional metrics but also developing specialized design to extract calibrated uncertainty signals.

A variety of uncertainty quantification techniques have been integrated with DL models to reduce labeling costs. In the simplest form, strategies originally developed for traditional shallow models, including LC, entropy, and PV, can be directly adapted to DNNs. For instance, these methods have been effectively utilized to filter high-entropy alloys~\cite{29}, predict phases across expansive composition spaces~\cite{53}, and identify highly relevant compounds while maintaining broad structural and chemical diversity~\cite{121,126}.
Despite the simplicity of the aforementioned methods, their reliance on raw model predictions makes them prone to the overconfidence issue inherent in DNNs, thereby limiting their reliability in DAL. To address this issue, ensemble-based methods have been introduced~\cite{17,21,42,107}, which estimate predictive uncertainty by aggregating the outputs of multiple models. For instance, ensembles of neural network interatomic potentials have been utilized to quantify uncertainty through the variance of force predictions across multiple model members~\cite{17}. Such formulations can be further refined with domain knowledge, such as utilizing weighted negative squared differences between independent networks to dynamically prioritize low-energy regions~\cite{107}.

Beyond statistical estimation from ensembles, alternative approaches have been developed to assess uncertainty from intrinsic or physical perspectives. Evidential DL offers a distinct perspective by combining the computational efficiency of traditional DNNs with the reliability of uncertainty estimation.
Unlike BNNs that rely on complex weight distributions, this approach retains the deterministic weights of standard networks but parametrizes the output as a higher-order probability distribution to model accumulated ``evidence'' derived from the inputs. This formulation enables calibrated, closed-form uncertainty estimation in a single forward pass. Crucially, this efficiency allows for the rapid prioritization of candidates in large-scale virtual screening and AL loops, achieving the high inference speed of traditional methods while avoiding the computationally expensive sampling required by ensembles. In molecular discovery, this approach is demonstrated by integrating evidential DL layers into standard architectures to predict continuous molecular properties~\cite{106}.
Similarly, calibrated gradient-based uncertainties, interpreted as model sensitivity to parameter perturbations, provide a cost-effective alternative for developing interatomic potentials with accuracy comparable to ensemble methods~\cite{90}.

\subsubsection{Deep distribution-based approach}
Unlike the traditional methods discussed above, which depend on fixed, pre-calculated descriptors (e.g., SOAP, ACSF) or static geometric constraints, distribution-based methods in DAL operate on dynamically evolving representations. In traditional settings, the feature representations are typically frozen in the input space; conversely, DAL leverages the strong representation learning capabilities of DNNs to construct a latent feature space that is jointly optimized with the task performance. In this context, the goal remains to effectively cover the chemical space to identifying representative samples within the model’s own learned semantic manifold. While the fundamental strategies, such as clustering and diversity sampling, remain conceptually similar, their application in DAL allows for a more flexible process where the definition of ``representativeness'' co-evolves with the model's understanding of the material landscape.

Clustering methods are also widely employed in distribution-based AL approaches, encompassing partition-based algorithms such as K-means and K-medoids, density-based algorithms like DBSCAN, and probabilistic clustering techniques~\cite{9,27}.
While K-means represents clusters by the average of data points, K-medoids utilizes actual data points as centers, a strategy that has been iteratively applied to explore vast candidate spaces involving 16 million catalysts~\cite{102}.
Such approaches frequently involve dimensionality reduction, such as Principal Component Analysis, to facilitate efficient clustering in low-dimensional feature spaces~\cite{82}.
In scenarios where the number of local minima is not predefined, Density-Based Spatial Clustering of Applications with Noise (DBSCAN) identifies candidate structures distant from stable configurations as noise. This approach avoids the formation of ill-defined clusters, thereby facilitating a more consistent mapping of the structural landscape~\cite{1}.

Another approach is to exploit the distance or similarity between data points in the feature space to avoid selecting redundant samples. Two primary AL strategies have been developed based on the reaction representation space, namely diversity sampling and adversarial sampling~\cite{123}. Diversity sampling selects unlabeled data that are least similar to the already labeled set to enhance model generalization. On the other hand, adversarial sampling is inspired by adversarial learning: small perturbations in a sample can cause incorrect predictions. Accordingly, this strategy prioritizes unlabeled samples whose predicted outcomes diverge significantly from the true labels of highly similar labeled data, thereby targeting regions where the model prediction is unstable.

\subsubsection{Hybrid deep approach}
Given the limitations of single-criterion AL methods, combining the multiple criteria discussed above has become a prevalent strategy for DAL. For example, considering both model uncertainty and feature characteristics in sample selection leads to a more comprehensive approach, aiming to select uncertain samples with reduced redundancy.

Specifically, a hybrid strategy has been developed to integrate uncertainty and diversity sampling across different optimization phases~\cite{19}. During the exploration phase, AL employs variance-based uncertainty sampling to select surfaces with high predictive uncertainty for computation. In the exploitation phase, BO is coupled with cosine similarity-based diversity sampling to prioritize diverse surfaces that are likely to exhibit high catalytic activity. Through iterative refinement, the model effectively identifies and optimizes highly efficient catalytic sites. Similarly, other frameworks combine structural dissimilarity with the predictive uncertainty of the potential function~\cite{103}. By integrating these two metrics, their method selects multiple structures without requiring frequent updates to the potential, thereby reducing redundancy.

\subsubsection{Others}
In addition to the mainstream methods discussed above, alternative AL approaches have been developed in the field of materials science to accelerate materials discovery.
A significant limitation of many aforementioned DAL methods is their reliance on pre-defined heuristics, which restricts their generalizability. To address this, a natural approach is to design selectors that can automatically learn querying strategies based on unlabeled data. Reinforcement learning (RL) represents a promising avenue for automating this process within DAL.
For instance, querying strategies can be optimized by combining fully Bayesian structured GP models with RL policy refinement to enable the co-navigation of hypothesis and experimental spaces~\cite{104}.
Similarly, multi-agent RL adaptive sampling strategies utilize complex reward functions to strategically select simulation restart points during the exploration phase~\cite{63}.
Beyond RL, genetic algorithms have also been incorporated into AL pipelines to generate candidate data. In this framework, generated candidates are selectively validated and incorporated via transfer learning to iteratively retrain the model~\cite{105}. This iterative process gradually expands the reliable prediction domain of the DNN toward regions containing materials with desired properties.

AL query strategies are diverse, employing various methodologies, models, and sampling techniques to efficiently select data for annotation. A comprehensive overview of these strategies, categorized by their underlying approaches and key components, is presented in Table~\ref{method_table}.

\begin{table*}[t!]
\caption{Summary of active learning methods for materials discovery. The strategies are divided into traditional and deep active learning approaches, providing a detailed summary of their methodologies, optimization goals, model architectures, and sampling criteria.}
\centering
\label{method_table}
\resizebox{1.0\linewidth}{!}{
\begin{tabular}{>{\centering\arraybackslash}m{2.5cm}
                |>{\centering\arraybackslash}m{3.0cm}
                |>{\centering\arraybackslash}m{5.6cm}
                |>{\centering\arraybackslash}m{4.8cm}
                |>{\centering\arraybackslash}m{4.5cm}}
\toprule
Category & Methodology & Goal & Model & Sampling Criteria \\
\hline

\multirow{9}[+70]{=}{\centering Traditional active learning}
& \multirow{2}[+10]{=}{\centering Bayesian optimization}
  & \multirow{2}[+12]{=}{\raggedright\arraybackslash Identifying the more promising candidates within vast chemical spaces by balancing exploration and exploitation}
  & \multirow{2}[+12]{*}{\centering Gaussian Process}
  & \multicolumn{1}{m{4.5cm}}{\raggedright\arraybackslash Exploitation-Exploration \cite{225,226,227,217,219,230,26,49,81,91,108,137,147,170,173}} \\
\hhline{~~~~-}
&   &
    &
    & \multicolumn{1}{m{4.5cm}}{\raggedright\arraybackslash Multi-Objective Bayesian Optimization (MOBO) \cite{57,3,89,176,99,12,179,175,124}} \\
\hhline{~----}

& \multirow{3}[+13]{=}{\centering Uncertainty-based approach}
  & \multirow{3}[+13]{=}{\raggedright\arraybackslash Prioritizing high-uncertainty samples to accelerate model convergence and efficient exploration}
  & Gaussian Process
  & \multicolumn{1}{m{4.5cm}}{\raggedright\arraybackslash Entropy~\cite{233}, Minimum confidence~\cite{95,206}, Predictive Variance \cite{4,25,30,33,38,84,93,96,110}} \\
\hhline{~~~--}
&   &
    & Gradient Boosting Regression, Support Vector Regression, Random Forest Regression
    &  \multicolumn{1}{m{4.5cm}}{\raggedright\arraybackslash Query by Committee \cite{111,28,50,64}} \\
\hhline{~~~--}
&   &
    & XGBoost,Gaussian Process, Gaussian Process Regression
    &  \multicolumn{1}{m{4.5cm}}{\raggedright\arraybackslash Others \cite{168,85,8,44,83,127,145}} \\
\hhline{~----}

& Distribution-based approach
  & \multicolumn{1}{m{5.6cm}|}{\raggedright\arraybackslash Understanding the underlying structure of the data and selecting batches of samples that best represent the unlabeled feature distribution}
  & Gaussian Process
  & \multicolumn{1}{m{4.5cm}}{\raggedright\arraybackslash Clustering-based \cite{232,72,114,58,72,76,172,35,130,135},
  Descriptor-based \cite{14,33,18,93,96,130}, Space-filling \cite{184,108,99,178,58,76,109,169}, Dynamic feature adjustment \cite{112,146}} \\
\hhline{~----}

& \multirow{1}{=}{\centering Utility-based approach}
  &\multicolumn{1}{m{5.6cm}|}{\raggedright\arraybackslash Identifying candidate samples that maximize the expected value of target properties for experimental validation}
  & Gaussian Process, Random Forest, XGBoost&\multicolumn{1}{m{4.5cm}}{\raggedright\arraybackslash Utility function \cite{169,181,5,183,119}} \\
\hhline{~----}

& \multirow{2}[+5]{=}{\centering Hybrid approach}
  & \multirow{2}[+5]{=}{Combining the strengths of multiple approaches to mitigate bias and sample redundancy}
  & \multirow{1}[+9]{=}{\centering Gaussian Process, Random Forest, Regression Tree}
  &  \multicolumn{1}{m{4.5cm}}{ Multi-stage \cite{172,76,58,72}} \\
  \hhline{~~~~-}
&
  &
  &
  &\multicolumn{1}{m{4.5cm}}{\raggedright\arraybackslash  Integrated hybrid approach \cite{56,166,10}} \\
\hline

\multirow{11}[+60]{=}{\centering Deep active learning}
& \multirow{3}[+28]{=}{\centering Deep Bayesian optimization}
  & \multirow{3}[+28]{=}{\raggedright\arraybackslash Optimizing the target objective of deep models with minimal labeled data by using uncertainty-aware Bayesian optimization in high-dimensional discontinuous, and non-stationary data}
  &  Gaussian Process, Convolutional Neural Network
  &  \multicolumn{1}{m{4.5cm}}{\raggedright\arraybackslash Deep kernel learning \cite{calandra2016manifold, wilson2016deep, wilson2016stochastic,47,ober2021promises}} \\
  \hhline{~~~--}

& & & Bayesian neural networks (BNNs)
  & \multicolumn{1}{m{4.5cm}}{\raggedright\arraybackslash Reliable Bayesian inference \cite{titterington2004bayesian, lampinen2001bayesian}, Monte Carlo Dropout \cite{gal2016dropout,2,92}, Others \cite{165,185}} \\
   \hhline{~~~--}

& & & Convolutional Neural Network, Gaussian Process
  &  \multicolumn{1}{m{4.5cm}}{\raggedright\arraybackslash EI strategy~\cite{47,92,120}, Dynamic Threshold Bayesian AL \cite{143},Multi-fidelity AL~\cite{163}, MOBO \cite{47}} \\
\hhline{~----}

& \multirow{3}[+13]{=}{\centering Deep uncertainty-based approach}
  & \multirow{3}[+12]{=}{Exploiting calibrated uncertainty signals from deep models to prioritize least reliable samples and guide exploration}
  & Multilayer Perceptron
  &  \multicolumn{1}{m{4.5cm}}{\raggedright\arraybackslash Traditional Uncertainty-based AL \cite{29,53,121,126}} \\
\hhline{~~~--}
&   &
    & Graph Neural Network, Multilayer Perceptron
    &  \multicolumn{1}{m{4.5cm}}{\raggedright\arraybackslash Ensemble-based methods \cite{17,21,42,107}} \\
\hhline{~~~--}
&   &
    & Embedded Atomic Neural Network
    &  \multicolumn{1}{m{4.5cm}}{\raggedright\arraybackslash Output-based metrics \cite{107}, Calibrated gradient-based uncertainties \cite{90}} \\
\hhline{~----}

& \multirow{2}{=}{\centering Deep distribution-based approach}
  & \multirow{2}{=}{Identifying representative samples within the model's own learned semantic space}
  & \multirow{2}[+5]{=}{\centering Artificial Neural Network}
  & \multicolumn{1}{m{4.5cm}}{\raggedright\arraybackslash Partition-based algorithms (K-means, K-medoids), Density-based algorithms (DBSCAN, probabilistic clustering techniques) \cite{9,27,82,102,1}} \\
\hhline{~~~~-}
&   &
    &
    & \multicolumn{1}{m{4.5cm}}{\raggedright\arraybackslash Distance and Similarity\cite{123}}  \\
\hhline{~----}

& Hybrid deep approach
  & \multicolumn{1}{m{5.6cm}|}{\raggedright\arraybackslash Combining multiple single-criterion strategies for DAL}
  & Graph Neural Network
  & \cite{19,103} \\
\hhline{~----}

& Others
  & \multicolumn{1}{m{5.6cm}|}{\raggedright\arraybackslash selectors that can automatically learn querying strategies based on unlabeled data}
  & Bayesian structured GP model
  & \cite{104,63,105} \\
\hline

\end{tabular}
}
\end{table*}

\section{Data foundations across regimes}\label{data}
To construct an efficient AL framework, it is not sufficient to design sampling strategies in isolation. It is also essential to systematically account for its tight coupling with both data scale and model capacity, thereby establishing a coherent and effective AL workflow. In this context, this chapter first surveys the diverse sources and acquisition routes of materials data, and discusses the initial data construction in the cold-start phase. We then examine, from the perspectives of ``small data and task-specific models''  and ``big data and foundation models'', highlighting how these regimes differ in the design of AL frameworks. Finally, the chapter elaborates on the dual role of AL in efficiently screening known chemical spaces and intelligently extrapolating into previously unexplored regions.

\subsection{Diverse data sources}
The importance of data in modern materials science has grown steadily, particularly in the process of materials discovery and design, where data-driven approaches have become central. In recent years, the rapid advancements in the field of materials science, especially driven by the accumulation of large-scale data, have significantly propelled the application of data-driven discovery methods. The vast amounts of data obtained from simulations and experiments provide a substantial foundation for materials discovery~\cite{agrawal2016perspective, butler2018machine}. Simulations, particularly first-principles methods, have generated large-scale computational databases, such as the Open Quantum Materials Database~\cite{kirklin2015open}, the Materials Project~\cite{jain2013commentary}, Automatic FLOW for Materials Discovery (AFLOW)~\cite{curtarolo2012aflow}, JARVIS-DFT~\cite{choudhary2020joint}, OC22~\cite{tran2023open}, OMat24~\cite{barrosoluque2024openmaterials2024omat24}, and MatPES~\cite{kaplan2025foundationalpotentialenergysurface}. In addition to these public databases, numerous studies have also generated customized computational data tailored to specific experimental tasks. These datasets enable the prediction of material properties prior to laboratory synthesis, thereby providing additional support for materials discovery~\cite{2,4,5,8,16,17,18,19,20,21,22}. While computational data provide important theoretical support for materials discovery, computational methods are inherently influenced by underlying models and assumptions, which may affect their applicability. Experimental data remain an indispensable component for validating physical and chemical properties across diverse material systems, especially for complex multi-scale and multi-physics phenomena. However, acquiring such data is often costly, time-consuming, and scarce.
To address this challenge, one effective approach involves literature mining and data extraction techniques~\cite{kononova2019text, 12, 29, 57, 94}, which systematically collect and integrate experimental data from existing research. The emergence of large language models (LLMs) and agents has further facilitated this process. These methodologies typically employ an LLM-centered workflow that first retrieves relevant document segments via semantic search or keyword filtering, then utilizes the LLM's reasoning capabilities to transform unstructured text into structured data formats, and finally incorporates iterative verification or external database cross-referencing to ensure the scientific accuracy of the extracted information~\cite{lai2023artificial, peng2025unlocking, ansari2024agent}. The diverse data sources utilized in materials discovery, ranging from computational repositories to literature-mined datasets, are synthesized in Table \ref{tab:data_sources}. These data provide a critical basis for data-driven AI modeling, underpinning the development of large-scale foundation models while facilitating specialized architectures for materials science applications.

\begin{table*}[htbp]
\caption{Summary of data sources and acquisition strategies in materials science. The datasets are  categorized into large-scale computational data, experimental data, and literature data.}
\label{tab:data_sources}
\centering
\resizebox{1.0\linewidth}{!}{
\begin{tabular}{
>{\centering\arraybackslash}m{2.2cm}
|>{\centering\arraybackslash}m{2.2cm}
|>{\centering\arraybackslash}m{3.0cm}
|>{\centering\arraybackslash}m{8.5cm}
|>{\centering\arraybackslash}m{2.0cm}}
\toprule
Category &  Source&Dataset&Description& References \\
\hline

\multirow{14}[+90]{=}{\centering  Computational data} &\multirow{14}[+90]{=}{\centering Multi-scale computational simulations , including DFT and MD}
& Open Quantum Materials Database&A comprehensive DFT dataset for inorganic crystalline materials, focusing on thermodynamic stability and structural properties for materials discovery
& \cite{kirklin2015open} \\
\hhline{~~---}

& & The Materials Project
& A massive repository of calculated properties for inorganic compounds, spanning known and theoretically predicted structures for battery and magnet research
& \cite{jain2013commentary} \\
\hhline{~~---}

& & AFLOWLIB
& An automated materials science database containing a vast amount of computational materials data, such as crystal structures, electronic structures, and thermodynamic properties
& \cite{curtarolo2012aflowlib} \\
\hhline{~~---}

& & JARVIS-DFT
& A diverse collection of 2D materials and bulk solids, providing high-accuracy DFT predictions for electronic, elastic, and topological properties
& \cite{choudhary2020joint} \\
\hhline{~~---}

& & OC20,OC22
& Large-scale datasets for electrocatalysts, focusing on surface-adsorbate interactions and reaction pathways for renewable energy applications
& \cite{chanussot2021open,tran2023open} \\
\hhline{~~---}

& & OMat24
& A vast dataset of inorganic bulk materials, featuring high structural and compositional diversity to improve the robustness of universal machine learning potentials
& \cite{barrosoluque2024openmaterials2024omat24} \\
\hhline{~~---}

& & MatPES
& A foundational potential energy surface (PES) dataset for diverse chemical environments, designed for training general-purpose interatomic potentials (UMLIPs)
& \cite{kaplan2025foundationalpotentialenergysurface} \\
\hhline{~~---}

& & MPF.2021
& A structured dataset for functional inorganic materials, specifically curated for predicting industrial properties in batteries and thermoelectrics.
& \cite{chen2022universal} \\
\hhline{~~---}

& & ANI-1x
& A high-precision dataset for organic molecules, focusing on molecular potential energy surfaces and conformational diversity for atomistic simulations
& \cite{smith2020ani} \\
\hhline{~~---}

& & Transition1x
& A specialized dataset for molecular chemical reactions, providing transition state geometries and barrier heights for reaction kinetics modeling
& \cite{schreiner2022transition1x} \\
\hhline{~~---}

& & MD17
& A high-fidelity dataset for small organic molecules, providing dense sampling trajectories for training machine learning force fields
& \cite{bowman2022md17} \\
\hhline{~~---}

& & Alexandria
& A large-scale collection of periodic crystalline materials, including stable and metastable phases for property prediction and stability analysis
& \cite{ghahremanpour2018alexandria} \\
\hhline{~~---}


& & Task-Specific Datasets
& Computational datasets constructed for specific tasks
& \cite{16,17,19,21,28,83} \\
\hline

\multirow{2}[+15]{=}{\centering Experimental data}
& \multirow{2}[+15]{=}{\centering Laboratory synthesis and measurements} &ICSD, CSD, COD&Gold-standard repositories for experimentally determined crystal structures (inorganic, organic, and metal-organic) for structural validation
& \cite{hellenbrandt2004inorganic,groom2016cambridge,gravzulis2009crystallography} \\
\hhline{~~---}

&  & Task-Specific Datasets &Specialized datasets for targeted material systems (e.g., polymers, superconductors), focusing on specific experimental performance metrics
& \cite{2,4,5,8,16,17,18,19,20,21,22} \\
\hline

\multirow{2}[+7]{=}{\centering Literature data}
& \multirow{2}[-6]{=}{\centering Traditional text mining, LLM-based framework} & MatScholar &A knowledge base extracted from millions of materials science papers, linking compositions to properties through NLP
& \cite{matscolar, lai2023artificial, peng2025unlocking} \\
\hhline{~~---}

&  & Task-Specific Datasets &Literature-mined data for specific material classes, capturing historical experimental results for composition design and discovery
& \cite{kononova2019text,12,29,57,94} \\
\hline

\end{tabular}
}
\end{table*}

\subsection{Cold-start strategy}
\label{cold}
To develop an effective data-driven AL method, it is essential to first determine which data should be used to train the initial model to start the iterative selection process, a challenge commonly referred to as the cold-start problem.
Some studies~\cite{chen2022making,hacohen2022active} reveal a striking paradox: at the initial selection stage, AL can perform no better, or potentially even worse, than random sampling. Addressing the cold-start problem is therefore crucial, as it can substantially affect performance in subsequent model training and sample selection cycles. Existing strategies for addressing the cold-start problem can generally be categorized into three approaches: random sampling, selecting initial samples guided by domain knowledge, and choosing representative samples based on feature distributions.

Random sampling strategies are frequently employed to address the cold-start problem by providing an unbiased initialization of the discovery cycle~\cite{20,60,100,142}. This approach typically involves the random selection of candidates from the configuration space to establish a representative baseline distribution of material properties. While computationally straightforward and devoid of heuristic bias, the efficacy of random sampling is often compromised in high-dimensional regimes. As the dimensionality of the descriptor space increases, the volume of the configuration space expands exponentially, causing random observations to become critically sparse and potentially failing to capture narrow yet critical regions of the chemical space within a constrained initial budget.

Regarding the domain-knowledge-guided approach, experts rely on their prior experience or existing literature to manually select samples.
A common practice involves the strategic selection of samples to ensure maximum coverage of prototypical coordination environments, functional groups, and diverse structural motifs, thereby anchoring the model with a robust prior understanding of the chemical diversity~\cite{54,58,2,75,102,116,85}.
Beyond de novo sampling, aggregating data from historical literature provides a viable alternative for model initialization. By leveraging insights from previous studies, this approach enables the model to establish a preliminary mapping of structure-property correlations, thereby grounding the subsequent discovery process in existing scientific knowledge~\cite{57}.
However, data aggregated from the literature may exhibit inconsistencies due to varying experimental or computational conditions. To mitigate these discrepancies, some studies involve the systematic recalculation of reported DFT data within a unified framework, ensuring a high-fidelity and consistent dataset for robust model training~\cite{67}.

Distinct from manual selection based on chemical intuition, diversity-driven sampling systematically explores the high-dimensional configuration space. Specifically, space-filling algorithms, such as LHS~\cite{99,124,184,178}, Graeco-Latin Squares~\cite{51}, or Sobol sequences~\cite{10,95,175}, are designed to achieve uniform coverage by partitioning the space into equiprobable regions, thereby minimizing unsampled voids.
In contrast to these grid-based designs, deterministic geometric approaches, including the Kennard-Stone algorithm~\cite{76} and farthest point sampling~\cite{58,109} iteratively select candidates to maximize inter-sample distances and minimize unsampled voids.
While the former two prioritize spatial spread, clustering-based approaches, such as K-means, represent another prevalent category for mitigating the cold-start problem by partitioning the inherent structure of the data distribution and selecting representative prototypes~\cite{72,114}.

\begin{figure}[h]
\centering
\includegraphics[width=1\linewidth]{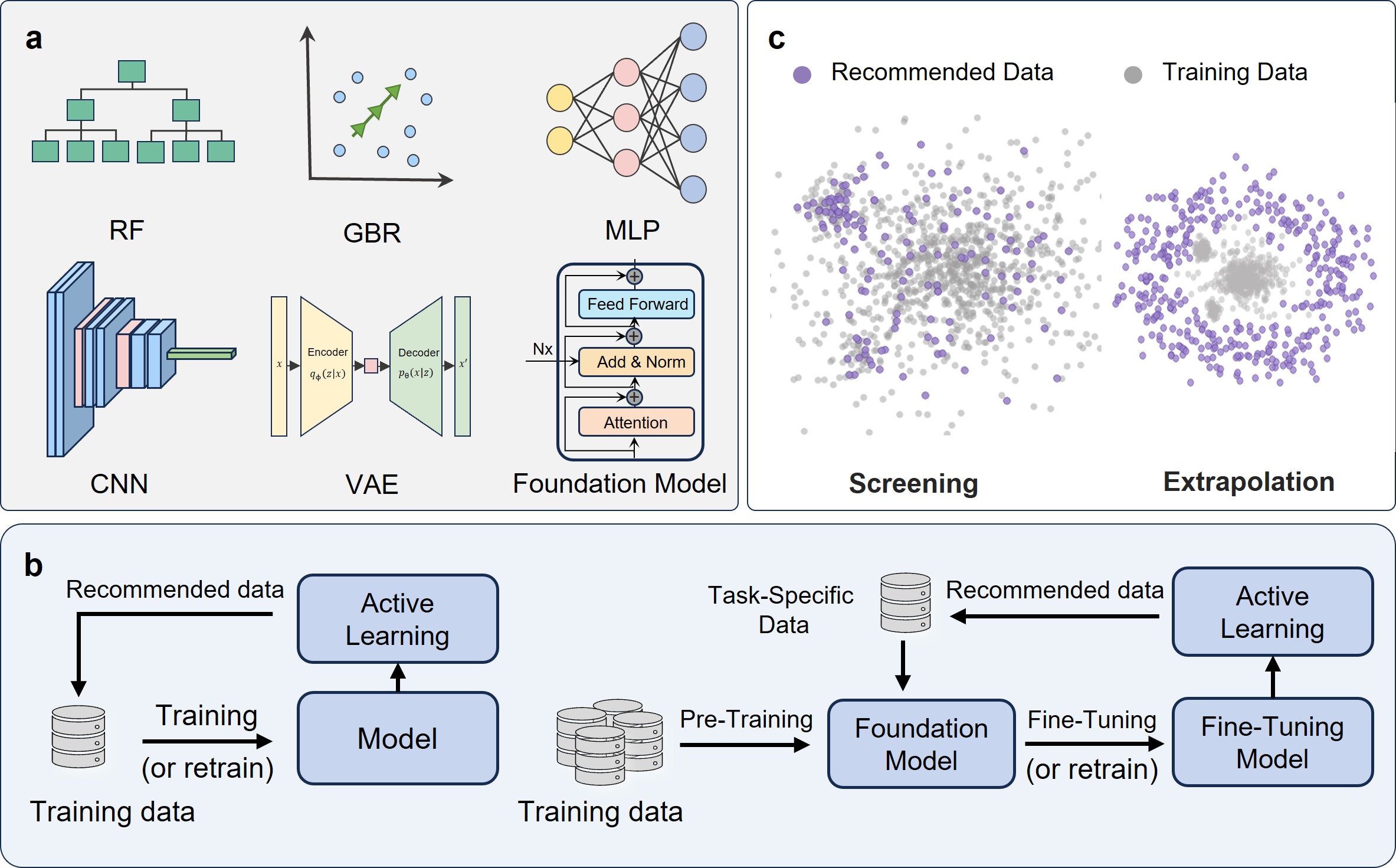}
\caption{\textbf{Data-driven paradigms and functional modalities of active learning in materials discovery.} \textbf{a} Representative machine learning architectures utilized as surrogate models, ranging from classical machine learning models to deep learning models and emerging foundation models. \textbf{b} Comparison of training workflows for materials discovery: a task-specific paradigm, often necessitated by small-data regimes, where specialized models are iteratively retrained on recommended data; and a foundation model paradigm, characterized by large-scale pre-training and task-specific fine-tuning within the AL loop to facilitate cross-task knowledge transfer. \textbf{c} Two core functional modes of AL in materials research. Screening focuses on mitigating data redundancy and identifying optimal candidates within known design spaces. Extrapolation prioritizes high-uncertainty samples to navigate unknown chemical regions, potentially guiding the discovery of novel materials beyond existing statistical boundaries. RF, random forest; GBR, gradient boosting regressor; MLP, multi-layer perceptron; CNN, convolutional neural network; VAE, variational autoencoder.}
\label{fig:data}
\end{figure}

\subsection{Small data and task-specific models}
Despite the abundance of public repositories, a vast majority of task-specific research remains severely data-constrained, typically relying on datasets comprising only dozens to hundreds of samples. This scarcity is fundamentally linked to the intrinsic heterogeneity and fragmentation of materials application scenarios. Existing large-scale data repositories often fail to cover the distribution required for specific tasks.
Consequently, a data-driven AL framework is essential to address task-specific data scarcity. Beyond the design of rigorous sampling strategies, the integration of appropriate surrogate models, comprehensive representation engineering, and data augmentation techniques is vital for overcoming small-data challenges.

Within the small-data regime of materials discovery, the choice of surrogate models is critical for effective AL. Traditional ML models, such as GPR and ensemble methods (e.g., RFs), are frequently employed to mitigate overfitting risks (Fig.~\ref{fig:data}a).
Among these, GPR stands out as the most prevalent surrogate model due to its unique probabilistic nature and ability to provide analytical uncertainty quantification. Unlike deterministic models that only yield point predictions, GPR treats target variables as distributions, outputting both predicted values and corresponding uncertainties for every point. This capability is essential for guiding acquisition functions such as EI and UCB, which balance exploration and exploitation based on these estimates~\cite{124,170,108,223,224,219}. Notably, prior studies have integrated domain knowledge, including human expertise, physical laws, and professional constraints, into kernel functions, enhancing physical feasibility and mitigating the risk of recommending candidates that appear optimal in silico but are experimentally infeasible~\cite{49,174,175}. Additionally, ensemble methods, such as RFs and XGBoost, are also widely adopted due to their ability to aggregate predictions from multiple models, thereby reducing prediction variance and improving stability~\cite{37,57,76,169}. Moreover, the diversity in predictions among ensemble members can be leveraged as an effective uncertainty measure in AL frameworks.

Furthermore, the efficacy of AL in small data regimes is often sensitive to descriptor design, particularly when employing traditional machine learning models. Since task-specific models lack sufficient data required to learn complex structural representations from scratch, effective descriptors are indispensable for ensuring model reliability and the overall performance of the AL loop. One prevalent approach involves the utilization of generalized descriptors, such as SOAP, which provide a formal framework to encode local atomic environments~\cite{14,18}. Alternatively, a more targeted strategy relies on hand-crafted, task-coupled features, such as electronegativity for catalysis or post-processing conditions for alloy synthesis~\cite{57,12}. The direct physical interpretability of these descriptors significantly reduces the learning burden on the surrogate model, allowing it to maintain robust predictive power when the available data points are extremely sparse.

Beyond structural representations, integrating generative models with AL offers a potential strategy to mitigate the sample scarcity in early iterations. The core strength of generative modeling lies in its capacity to capture the underlying probability distribution of the training dataset and subsequently sample from this learned distribution to produce novel outputs. By augmenting the existing dataset with synthesized real or virtual candidate materials, the AL framework can perform more robust exploratory sampling. The implementation of this generative strategy primarily relies on two distinct architectural frameworks: Variational Autoencoders (VAEs) and Generative Adversarial Networks (GANs). VAEs facilitate smooth interpolation within the chemical space by mapping complex material data into a constrained, continuous latent space, ensuring that synthetic samples remain statistically faithful to the established data distribution. A systematic evaluation of diverse generative techniques highlighted the superior capacity of Tabular VAEs in maintaining high distributional fidelity~\cite{bao2025bio}. In catalytic research, such augmentation has refined model decision boundaries, enabling the achievement of an $R^2$ of 0.96 for $\mathrm{CO}_2$ methanation performance prediction.
In contrast, GANs achieve high-fidelity synthesis through an adversarial game-theoretic objective. This architecture comprises two competing components: a generator that captures the underlying data distribution to synthesize fake samples, and a discriminator that distinguishes between synthetic and real data. Through this continuous adversarial training, the generator learns to produce high-quality samples that are indistinguishable from real experimental measurements. By leveraging Wasserstein GAN synthesized samples to expand the sparse training set, researchers have achieved the rapid identification of optimized components, such as reducing oxide layer thickness by 17\% in a single iteration, thereby resolving small-sample optimization bottlenecks through high-resolution data synthesis~\cite{WANG2025182634}.

Although data generation strategies based on generative models have achieved notable success across a range of systems, the reliable generation of high-quality data that are both physically consistent and experimentally feasible remains a central challenge. On the one hand, the generative process must strictly respect physical constraints, chemical stability, and the synthesizability of molecules or materials; otherwise, model exploration may be driven into experimentally unverifiable regions of the design space. On the other hand, under small-data regimes, existing generative models suffer from training-data bias, limiting their ability to preserve distributional fidelity while effectively sampling high-performance or rare regions.
Overall, the judicious integration of surrogate models, descriptor design, and AL strategies are pivotal for task-specific applications. This synergy not only minimizes experimental and computational expenditures but also accelerates the systematic exploration of underlying material mechanisms and expansive design spaces.

\subsection{Big data and foundation models}
Parallel to the development of task-specific models, the continuous aggregation of high-throughput computational and experimental data has fostered the emergence of general-purpose datasets and foundation models. By leveraging massive repositories such as the Materials Project and AFLOW, these models surpass the limitations of manual feature engineering, learning sophisticated latent representations that capture the universal physical and chemical principles.
Within the AL framework, such foundation models may provide a robust, general prior that can directly guide sampling. Furthermore, they can be adapted to specific discovery tasks through active fine-tuning, thereby further improving their predictive accuracy (Fig. \ref{fig:data}b).

Using pre-trained models as surrogate models to guide AL leverages their profound prior knowledge and universal token representations to effectively mitigate the prediction challenges in the cold-start phase and the heavy reliance on complex feature engineering faced by traditional models. For instance, a study proposed the LLM-AL framework, which transforms experimental parameters into text prompts and utilizes the reasoning capabilities of Claude-3-Sonnet to select high-value samples without fine-tuning; this approach reached optima across four diverse materials datasets using less than 30\% of the data, outperforming traditional models by over 70\%. This closed-loop AL paradigm demonstrates that pre-trained LLMs can serve as training-free, interpretable surrogate models for cross-domain autonomous materials discovery~\cite{wang2025training}.
Beyond the direct utilization of pre-trained foundation models for zero-shot sample selection, the ``Pre-training + Active Fine-tuning'' paradigm is emerging as a strategic approach to mitigate the scarcity of high-fidelity data in specialized material domains. Within this framework, AL algorithms are employed to identify the most informative data points based on the model’s generalized physical priors, which are subsequently used for incremental fine-tuning. This strategy aims to bridge the gap between universal physical representations and task-specific energy landscapes or property distributions, potentially enhancing both predictive reliability and convergence efficiency in niche chemical spaces~\cite{kang2024get,masood2025molecular}. Notably, distribution shifts may exist between task-specific data and the pre-training data distribution, which may require appropriate selection strategies to better adapt to out-of-distribution data.

In addition to using general-purpose foundation models to guide AL sampling, AL in turn facilitates the construction of the massive datasets on which these models depend. It is achieved by intelligent sampling strategies that eliminate redundancy, maximize information gain, and systematically broaden the exploration of chemical space.
In DeepMind’s GNoME study, AL functioned as a ``data flywheel'', identifying 2.2 million stable crystals from 220 million candidates and expanding known material databases by an order of magnitude. Beyond identifying stable candidates, this iterative process captured extensive ionic relaxation trajectories, providing the diverse non-equilibrium data necessary to pre-train universal interatomic potentials. Consequently, these pre-trained models exhibit exceptional zero-shot generalization, enabling the efficient screening of novel candidates across uncharted chemical spaces~\cite{merchant2023scaling}.
Similarly, the SurFF framework leverages a data-efficient AL strategy to address the challenge of predicting surface exposure across vast intermetallic crystal spaces. By employing an uncertainty-based and diversity-based acquisition function, SurFF iteratively selected the most informative surfaces from a design space of over 284,000 candidates for DFT calculations. This process resulted in a comprehensive database of 12,553 unique surfaces and 344,200 single points, enabling the development of a foundation model that achieves DFT-level precision in surface energy prediction and morphology determination with a $10^5$-fold acceleration~\cite{yin2025surff}.
Despite these milestones, the systematic deployment of AL for generalized datasets remains in its nascent stages and faces persistent challenges such as multi-scale modeling and multi-modal alignment. Future research should therefore transcend simple gains in sampling efficiency and focus on developing physically robust, universal frameworks where AL and foundation models co-evolve.

\subsection{Screening and extrapolation}
When designing an AL framework for data-driven materials discovery, it is essential to consider not only the data regime but also the distributional gap between existing and target data. In this context, AL not only enables efficient screening of candidates from an existing data pool, but also supports extrapolation into unexplored regions of chemical space, thereby expanding the boundaries of materials knowledge. (Fig. \ref{fig:data}c).

Data screening is widely utilized to extract representative and diverse subsets from expansive data distributions, ensuring that the most informative samples are prioritized for model training. Systematic evaluations of major repositories such as JARVIS, the Materials Project, and the Open Quantum Materials Database reveal that data redundancy can reach as high as 95\%~\cite{20}. This pervasive redundancy implies that only 5\% of the data need to be retained for training, with a minimal impact on the model's predictive performance within the distribution. Redundant data not only fails to improve the model's performance on out-of-distribution samples but may, due to distributional biases, exacerbate its generalization deficiencies, leading to severe performance degradation in real-world materials discovery scenarios.
Therefore, utilizing AL to select representative subsets for annotation is of significant importance, as it substantially reduces training costs while preserving model performance. More broadly, this paradigm reflects a shift from the mere pursuit of ``data quantity'' to the optimization of ``data informativeness'' in both dataset construction and model training.

In addition to reducing data redundancy, AL serves as a strategic navigator within the known materials space, significantly alleviating the computational and experimental burden associated with exhaustive screening~\cite{8,17,19,28,47,54,56}. Given the highly non-linear nature of structure-property relationships, AL aims to isolate the ``critical minority'' of candidates from the vast design space. By intelligently prioritizing these high-value regions, AL effectively bypasses inefficient trial-and-error, ensuring that every calculation or synthesis step efficiently guides the search toward optimal solutions.
For instance, in catalyst discovery, guided closed‑loop optimization has identified high‑performance compositions with far fewer evaluations—often below 5\% of the initial candidate pool—while achieving significant performance enhancements, such as a five‑fold increase in alcohol synthesis rate or over 80\% Faradaic efficiency for ethylene production~\cite{5,3}. Its feasibility has been further demonstrated across other pivotal domains of materials research, including alloy design, organic molecule screening, and synthesis process optimization~\cite{8,100,120,3}.

Screening-oriented AL approaches that are primarily efficient for local optimization fundamentally limit the identification of materials with novel compositions or structures outside the existing data distribution.
However, one challenge of materials discovery lies in capturing meaningful outliers that extrapolate beyond established knowledge boundaries (Fig. \ref{fig:data}c). By directing exploration toward regions characterized by sparse data distributions and high uncertainty, AL identifies promising candidates in previously unexplored chemical spaces.
A key strategy for enabling extrapolation is the prioritization of candidates with high predictive uncertainty. By evaluating the posterior variance from GPR or the predictive discrepancy within model ensembles, AL identifies target regions that lie beyond the established data distribution~\cite{8,111}.
Furthermore, the D-optimality criterion established a rigorous framework for identifying extrapolative regions by maximizing the determinant of the design matrix. In practice, the MaxVol algorithm translates this criterion into an efficient evaluation mechanism via the extrapolation grade ($\gamma$), which pinpoints candidates residing at the periphery of the feature space. By identifying these high-volume sub-matrices, the approach enables the autonomous discovery of critical configurations missing from the training set, effectively guiding the expansion of the knowledge boundary toward meaningful outliers~\cite{86,88,178}.
Beyond static batch selection, AL can also actively steer dynamic exploration by incorporating uncertainty directly into the physical potential. A representative case involves the configurational exploration of alanine dipeptide and MIL-53(Al). By subtracting energy uncertainty from the predicted potential energy surface, this approach constructs an altered energy landscape that generates bias forces, driving the simulation trajectories directly into extrapolative regions and rare events typically missed by conventional MD. This mechanism facilitates the discovery of meaningful outliers and achieves uniform model accuracy at a lower computational cost than traditional ensemble-based methods~\cite{90}.

The integration of generative models with AL represents a pivotal strategy for achieving extrapolation. By implementing objective-driven guidance criteria, AL imposes functional conditioning on the generative process. This shifts the sampling trajectory from mere distribution imitation to targeted evolution within a defined parameter space. This guidance mechanism ensures that the generative process transcends the constraints of interpolation, enabling the discovery of physically meaningful novel configurations within data-sparse regions along property frontiers. In the design of ultrahigh thermal conductivity materials, the on-the-fly training of machine learning potentials via AL effectively addresses the energetic deviations of generated structures, thereby steering the sampling trajectory from random exploration toward thermodynamically stable, high-performance regimes to identify 34 carbon polymorphs with predicted thermal conductivities exceeding those of most known materials~\cite{160}. Similarly, in the inverse design of energetic molecules, AL utilizes iterative feedback on molecular stability to enhance the generalization of property predictors across unseen chemical spaces, shifting the generative sampling distribution toward high-performance boundaries that lie multiple standard deviations beyond the original training data range~\cite{antoniuk2025activelearningenablesextrapolation}. However, developing generation frameworks that jointly balance data quality, physical consistency, exploration efficiency, and experimental verifiability constitutes a key challenge for translating these approaches from proof-of-concept studies to practical materials discovery.

\section{Accelerating the materials research pipeline}\label{application}
\begin{figure}[h]
\centering
\includegraphics[width=1\linewidth]{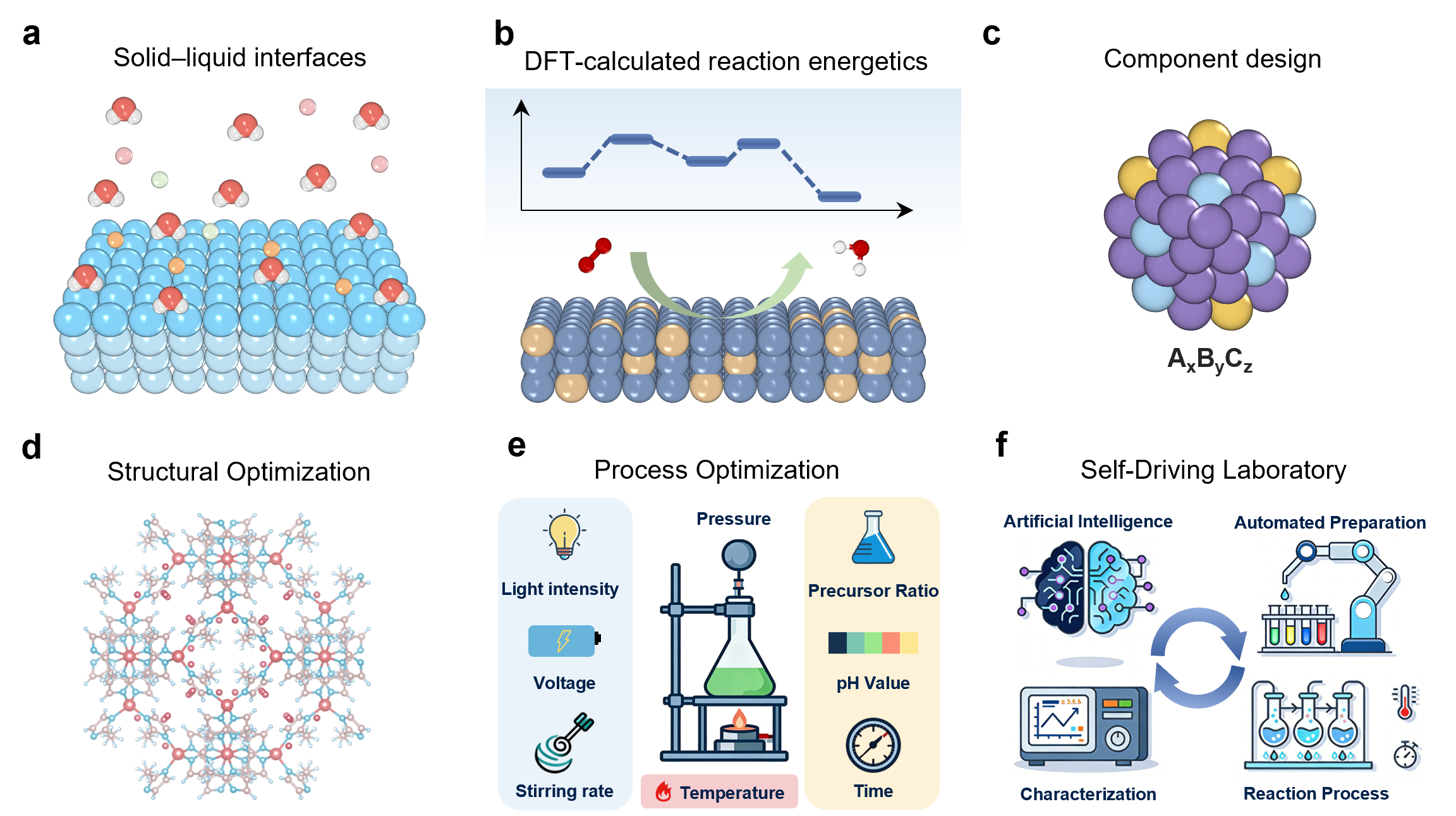}
\caption{\textbf{Applications}. \textbf{a} Interfacial reaction dynamics simulations. \textbf{b} DFT-calculated Gibbs free energy profiles for the electrocatalytic reaction pathway. \textbf{c} Compositional design of metal nanoparticles. \textbf{d} Structural optimization of metal-organic frameworks. \textbf{e} Optimization of synthesis process parameters. \textbf{f} Autonomous closed-loop discovery in a self-driving laboratory.} \label{fig:figApplication}
\end{figure}

By integrating sampling strategies, data distributions, and model capacity within a unified framework, AL serves as a systematic framework for accelerating the materials research pipeline. This section examines how these strategies enhance key stages of materials discovery, as illustrated in Fig. \ref{fig:figApplication}. We review the deployment of AL across computational simulation, compositional and structural design, and process optimization, advances that are progressively converging toward closed-loop, self-driving laboratories (SDLs). Notably, while AL has been widely adopted in materials discovery and synthesis, its application in characterization remains an emerging area, pointing to a critical opportunity for future development.

\subsection{Enhanced computational simulations}
Computational simulations, encompassing MD simulations (Fig. \ref{fig:figApplication}a) and first-principles calculations (Fig. \ref{fig:figApplication}b), are essential for probing atomic-scale structures and processes. While first-principles calculations offer ab initio-level accuracy for electronic structures, they are typically static, confined to zero-Kelvin ground states and thus unable to capture thermodynamic fluctuations or kinetic evolution. Conversely, MD simulations explicitly capture temporal evolution and finite-temperature fluctuations, thereby enabling the description of complex dynamical phenomena. However, classical MD simulations often suffer from a lack of chemical accuracy due to their reliance on classical potentials. Although AIMD overcomes this by treating electronic structure explicitly, the associated computational burden renders simulations across extended spatiotemporal scales intractable.

Machine learning potentials (MLPs) have emerged as a key and potentially approach to mitigating this long-standing trade-off between accuracy and scale~\cite{tang2025deep}. By using ML to predict energies and forces with near-DFT fidelity, MLPs bridge the gap between ab initio accuracy and large-scale molecular dynamics, enabling simulations of complex systems over relevant time and length-scales. The development of high-accuracy MLPs, however, traditionally depends on large, expensively labeled datasets from first-principles calculations. AL achieves high data efficiency by strategically guiding the selection of the most informative data within vast configurational spaces. This enables models to reach target accuracy with minimal data while simultaneously steering the exploration toward specific scientific objectives, such as locating global energy minima in crystal structures, capturing high-energy transition states for rare events, or resolving elusive phase boundaries in multicomponent alloys.

In practice, AL is seamlessly embedded into the MD simulation or structural exploration loop, enabling concurrent model training and system discovery through a ``simulate-verify-update'' closed-loop strategy. As data arrive in a continuous stream during the simulation, the system autonomously evaluates the informational value of each frame to determine whether a high-fidelity first-principles calculation is required. Once triggered, these data points are integrated into the training set to retrain the model, effectively reinforcing the potential energy surface in real time. This iterative refinement ensures that the MLP adaptively evolves to maintain robust accuracy precisely where the simulation requires it most.

To navigate the combinatorial explosion of atomic arrangements within expansive structural search spaces, AL-driven MLPs offer a systematic framework for the rapid identification of stable structures, serving as a foundational step in materials discovery. Early work demonstrated that AL could efficiently guide the relaxation of randomly generated crystal candidates toward local energy minima using neural network potentials~\cite{1}. This paradigm has evolved into integrated workflows that combine generative structural proposals with physics-based validation. For example, generative models were used to propose diverse candidate crystals whose stability was then evaluated using interatomic potentials refined through AL. In one implementation, such an iterative, uncertainty-aware strategy enabled the targeted screening of carbon polymorphs for high lattice thermal conductivity, effectively narrowing a broad candidate space to a focused set of promising materials~\cite{160}. Beyond generative sampling strategies, physics-aware approaches can effectively overcome the limitations of static training datasets. A temperature-driven AL protocol was developed to automatically capture critical configurations across diverse thermodynamic states by iteratively refining the potential during molecular dynamics runs performed at progressively higher temperatures~\cite{41}. Moving beyond the identification of static minima, AL strategies have been increasingly adapted to resolve the complex kinetic pathways and functional responses of materials under diverse conditions. In the study of ductile inorganic materials, an automated workflow integrated AL with the climbing-image nudged elastic band method to efficiently reconstruct the interlayer slip potential energy surface and systematically determine the minimum-energy pathway for static deformations. Building on these foundational capabilities in structural and energetic mapping, the application of AL has expanded into more complex temporal and thermodynamic regimes, providing a robust framework for investigating phase transitions, long-range transport phenomena, and Reactive MD.

Simulating phase transitions such as melting, crystallization, and amorphization requires capturing the collective rearrangement of thousands of atoms across complex kinetic pathways. A fundamental bottleneck arises because crystal nucleation necessitates a specific critical volume that typically exceeds the scales manageable by traditional ab initio molecular dynamics, which is often limited to hundreds of atoms and prone to periodic boundary artifacts. While MLPs significantly alleviate these spatiotemporal constraints, conventional training strategies predominantly focus on near-equilibrium stable states. Consequently, they often lack the necessary coverage of rare events and high-energy transition states that are indispensable for accurately resolving nucleation barriers. AL facilitates a more targeted exploration of the configurational landscape by identifying high-uncertainty regions associated with rare events, thereby enhancing the representativeness of the potential energy surface across diverse phase space.
An adaptive Bayesian inference framework, known as FLARE, captures rare events through the introduction of a fully interpretable and low-dimensional force field model. By rigorously partitioning epistemic and aleatoric uncertainties, this approach identifies high-energy configurations to trigger first-principles calculations only when necessary, thereby maintaining chemical accuracy with minimal computational overhead~\cite{33}. As research shifts toward complex environments, Bayesian AL has efficiently reproduced high-pressure behavior in silicon carbide~\cite{30}, captured ripple-driven nucleation in 2D stanene~\cite{110}, and surmounted trajectory constraints to simulate the $\mathrm{HfO_2}$ melt-quench process~\cite{135}. To bridge the gap between microscopic trajectories and macroscopic thermodynamics, recent strategies have moved beyond purely data-driven potentials by leveraging AL to parameterize physically motivated effective Hamiltonians. By integrating Bayesian inference with these physics-informed models, this hybrid approach offers a decisive efficiency advantage over brute-force simulations, enabling the precise prediction of ferroelectric phase transition temperatures ($T_c$) in super-large-scale systems containing over $10^7$ atoms~\cite{142}. Building on such macroscopic predictions, this paradigm can be further extended to the high-throughput mapping of phase diagrams across broader chemical spaces. By embedding Gibbs' phase rule directly into the learning loop to prune the search space and augment training data in multiphase regions, this approach resolves intricate ternary landscapes with an eight-fold reduction in experimental overhead, transforming the mapping of material stability into a targeted, intelligence-led discovery process~\cite{127}.

The investigation of material dynamic behavior is fundamentally constrained by a temporal bottleneck. The convergence of non-equilibrium transport properties, such as ion diffusion and viscous flow, necessitates nanosecond-to-microsecond trajectories that exceed the limits of traditional ab initio molecular dynamics. AL has emerged as a strategic solution to bridge this divide by enabling the construction of MLPs that maintain quantum-mechanical fidelity over extended timescales.
For complex systems lacking reliable empirical force fields, an effective paradigm utilizes inexpensive classical potentials for initial sampling, followed by AL to isolate a sparse set of representative configurations. This framework achieves a 19,000-fold acceleration relative to density functional theory, providing the nanosecond-scale sampling depth required to converge ionic conductivity in molten salts~\cite{130}.
Similar methodologies extend to the domain of spin dynamics, where an AL-driven framework enables the prediction of spin-lattice relaxation times in open-shell coordination compounds with an 80\% reduction in computational overhead while navigating complex magneto-structural correlations~\cite{161}.
Conversely, for many molecular systems such as liquid electrolytes, while classical force fields often provide a physically robust formalism, the parameterization within high-dimensional spaces remains a formidable challenge. In these contexts, the role of AL shifts from constructing new potentials to the intelligent optimization of existing parameters. For example, a hybrid strategy combining genetic algorithms with GPR was deployed to efficiently navigate the high-dimensional parameter space of sulfone electrolytes. Rather than rebuilding the potential energy surface from scratch, this approach rectified the severe viscosity artifacts inherent to generic force fields using fewer than 300 reference calculations~\cite{159}.

Reactive MD requires a flexible model of the PES that is chemically accurate in describing bond breaking and formation, thereby systematically unraveling the transient mechanisms that govern reaction rates. While MLPs achieve quantum-mechanical fidelity in describing fundamental molecular interactions~\cite{131},``insufficient sampling" remains a primary bottleneck in simulating complex chemical processes. Traditional models often approximate catalytic surfaces as static entities, thereby overlooking the dynamic reconstructions essential to reactive environments. Uncertainty-driven AL addresses this limitation by providing a general methodology for constructing accurate potential energy surfaces for gas-solid reactions~\cite{107}. This paradigm has been applied to platinum catalysts under varying hydrogen coverage, where explicit simulation of the gas-solid interface accurately captured both surface and bulk dynamics, as well as surface and subsurface hydrogen diffusion~\cite{4}.
However, the dynamic nature of catalysts extends beyond local thermal fluctuations of surface atoms, fundamentally manifesting as global structural reconstructions induced by the reaction environment. Furthermore, the introduction of topology-guided sampling has enabled the identification of metastable active phases, such as $\mathrm{PdH}_x$ hydrides and $\mathrm{PtO}_x$ oxides~\cite{144}. Similarly, an automated training protocol for $\mathrm{CO}_2$ hydrogenation over $\mathrm{In}_2\mathrm{O}_3$ revealed that the in situ generated, reduced top-layer surface serves as the genuine active center~\cite{96}. The introduction of explicit solvents induces an exponential increase in system complexity, particularly at solid-liquid interfaces. To address this, specialized AL strategies have been developed to isolate and sample critical reactive regions, enabling the tracking of large-scale silicate polymerization dynamics involving continuous multi-step bonding events within complex aqueous environments.
Furthermore, by leveraging transferability strategies tailored for liquid environments, researchers have achieved precise simulation of nanosecond-scale solvent dynamics while resolving model generalization challenges across systems ranging from simple aqueous solutions to complex organic electrolytes~\cite{18,29}.

\subsection{Compositional design optimization}
Compositional design is one of the most direct and effective strategies to control material properties. By systematically adjusting the types and ratios of elements within materials, researchers can significantly modulate their electronic structure, crystal stability, interface properties, and ultimately their functional performance (Fig. \ref{fig:figApplication}c). Especially in key application areas such as energy, catalysis, and energy storage, rational compositional design not only determines the performance limits of materials but also directly impacts their sustainability and cost feasibility. However, as compositional complexity increases, the combination space grows exponentially, making traditional exhaustive or experimental methods inefficient. To address this, AL has been widely applied in materials composition optimization research in recent years~\cite{8,116,111,113,9,134,120,92}. This strategy builds a feedback loop between model prediction and experimental validation, enabling the intelligent selection of the most representative candidate compositions from vast combination spaces, thereby accelerating the material screening and discovery process. Whether optimizing molar ratios in high-entropy oxides for hydrogen production~\cite{169} or tuning perovskite oxides~\cite{9} and diamond-like compounds~\cite{111}, AL demonstrates a universal capability to enhance model extrapolation and accelerate discovery across diverse material classes.

Alloys represent a quintessential example of compositionally tunable materials. By incorporating two or more metallic elements, they enable atomic-scale electronic structure reconstruction, geometric structure regulation, and synergistic effects, resulting in performance advantages unattainable by a single element.
In electrocatalysis, considerable attention has been paid to improving the activity, stability, and resource accessibility of catalysts by adjusting the component ratios in alloys. Initial efforts prioritized rapid screening across expansive binary and multicomponent landscapes to establish atomistic structure-performance relationships. By deploying AL to predict adsorption energies across thousands of binary alloy surfaces, researchers successfully identified high-performance candidates for the hydrogen evolution reaction and $\mathrm{CO_2}$ reduction reaction ($\mathrm{CO_2RR}$) with properties approaching theoretical optima, thereby drastically reducing the experimental search space~\cite{8}. A similar approach has been extended to explore novel complex material systems, such as high-entropy alloys, and to optimize the cost-effectiveness of traditional precious-metal catalysts, demonstrating the capability of AL for multi-dimensional optimization in materials discovery~\cite{116,75,60}. While catalytic activity remains a primary metric, it is seldom the sole determinant of a material’s viability in practical applications, where stability, cost-effectiveness, and selectivity are equally paramount. To navigate the inherent trade-offs between competing performance requirements, Pareto-based AL frameworks have demonstrated exceptional utility, particularly in the design of bifunctional catalysts. Within the $\mathrm{Ni\text{-}Fe\text{-}Co}$ ternary space, this paradigm has enabled the simultaneous optimization of anodic and cathodic efficiencies, achieving a synergistic performance balance that traditional screening often fails to capture~\cite{100}. The robustness of this multi-objective optimization approach is further evidenced by its application to other challenging electrochemical and chemical systems. For instance, it has been employed to mitigate the inherent tension between anodic efficiency and discharge voltage in Mg-air batteries, as well as to balance reaction rates with olefin selectivity in alkyne semi-hydrogenation~\cite{57,2}. The scope of AL-driven discovery has recently expanded from bulk metallic stoichiometry to the precise orchestration of metal-ligand coordination. For instance, AL strategies have been employed to navigate a $\mathrm{Mn\text{-}Fe}$ catalyst space of nearly 16 million structures, identifying optimal combinations of axial and planar ligands that balance methane hydrogen atom transfer with methanol release~\cite{102}. Similarly, integrating AL with graph neural networks has facilitated the rapid screening of dual-atom catalysts, identifying 3$d$ transition metal pairs (e.g., $\mathrm{Co\text{-}Fe}$ and $\mathrm{Co\text{-}Zn}$) whose predicted oxygen electrode activities align closely with experimental benchmarks~\cite{19}.

Alloy design not only plays a crucial role in catalysis but also enhances the application potential of materials in high-temperature, corrosion-resistant, and functional environments by regulating thermal expansion, structural stability, formation energy, and mechanical properties. For example, an AL framework integrating generative models with two-stage ensemble regression has been developed to map high-dimensional compositional spaces into a lower-dimensional latent space, incorporating physical descriptors. Using only 17 experiments, this strategy identified two new Invar high-entropy alloys with an ultralow thermal expansion coefficient of $2 \times 10^{-6}~\mathrm{K^{-1}}$, matching the performance of conventional binary Invars~\cite{29}. Similarly, graph convolutional networks combined with AL have been employed to predict the formation energies of $\mathrm{Pd\text{-}Pt\text{-}Sn}$ ternary alloys and efficiently screen for stable structures~\cite{78}. Furthermore, a multi-objective Bayesian AL strategy, which incorporates data uncertainty, enables a balance between strength and ductility; this strategy has been successfully applied to the discovery of novel lead-free solder alloys~\cite{26,91}.

In molecular design, precise and programmable editing of functional groups enables targeted property tuning, such as optimizing electrochemical performance in redox flow batteries through systematic substitution on molecular backbones~\cite{134}. This approach also facilitates the efficient discovery of photosensitizers by targeting the singlet-triplet energy gap ($\Delta E_{\mathrm{ST}}$) as the key photophysical descriptor within a self-improving Bayesian AL loop, enabling directed exploration of a chemical space exceeding 7 million molecules with minimal first-principles data~\cite{120}. Moreover, in the design of composites and heterogeneous systems, an inverse design framework enables the direct mapping of preset mechanical targets to the spatial distribution of materials. By accurately identifying micro-topologies with exceptional toughness using minimal data, this approach opens new possibilities for the efficient creation of materials across multiple scales~\cite{113}.

\subsection{Structural optimization}
Parallel to compositional tuning, structural design offers a way to modulate material properties by focusing on spatial arrangements and geometric complexity. In catalysis, for instance, key properties such as reaction energy barriers, selectivity, and efficiency are predominantly dictated by surface structure, atomic arrangement, and crystal morphology. While the rational optimization of these structural features is paramount, the vast and complex configuration space renders conventional exhaustive or random screening approaches prohibitively expensive. By effectively balancing computational cost with search accuracy, AL has been implemented across diverse material regimes, ranging from the atomic-scale engineering of metallic and oxide surfaces to the topological regulation of complex porous and organic systems (Fig. \ref{fig:figApplication}d).

In well-defined metallic systems, the research focus has evolved from static configurational sampling. For instance, AL frameworks have been utilized to identify optimal co-adsorption configurations on metallic surfaces, such as $NH*_OH*$ and $N*_NO*$ on $\mathrm{Pt(111)@Cu_2}$, using less than 10\% of the data required by exhaustive searches~\cite{35}. Beyond static arrangements, the integration of AL with Graph Neural Networks enables the systematic geometry optimization of transition states and intermediates, facilitating a deeper mechanistic understanding of pathways like $\mathrm{CO_2}$ reduction and C-C coupling on Cu-based alloys~\cite{143}. As the catalytic environment shifts toward oxides, the increased electronic and bonding complexity often necessitates the incorporation of physical principles into AL workflows. Physics-informed models, such as those utilizing rotationally invariant principles, have achieved adsorption energy predictions across diverse oxide structures with reduced reliance on expensive DFT calculations~\cite{42}. Furthermore, AL-driven screening has been applied to search for new zeolite structures with high shear moduli~\cite{67}.

The design of porous and heterogeneous materials emphasizes the synergistic regulation of topology and interfacial functionalization. In metal-organic frameworks, AL has been employed to decode the interplay between geometric factors, such as pore size, and material stability or diffusion coefficients, thereby leading to the identification of high-permeability materials~\cite{122,50}. Similarly, interface engineering in nanomaterials and 2D heterostructures benefits from active screening. Recent applications include the optimization of surface ligands for perovskite nanocrystals to enhance quantum yield~\cite{54}, and the navigation of high-dimensional stacking landscapes in transition metal dichalcogenides to tune band gaps and thermoelectric properties~\cite{22}.
Finally, AL facilitates molecular engineering in the vast chemical space of organic and hybrid systems. At the molecular level, this strategy enables the rapid identification of semiconductors with high charge transport capabilities~\cite{108} and the high-precision prediction of excited states in aromatic hydrocarbons using minimal labeled samples~\cite{77}. The utility of AL extends to the regulation of microscopic textures and hybrid interfaces, where it assists in designing polycrystalline microstructures with specific plastic anisotropy~\cite{38}.

\subsection{Process parameter optimization}
Process optimization primarily involves understanding process-property correlations to find the optimal parameter condition, thereby yielding materials with superior properties.
Material optimization typically involves multiple key parameters, such as temperature, pressure, reaction time, and solvent concentration (Fig. \ref{fig:figApplication}e), and the relationships between these parameters are highly nonlinear and interwoven.
Moreover, in multi-stage processes, the optimization at one stage can impact subsequent steps, making it difficult to achieve global optimization through single-stage methods.
Furthermore, due to the lack of real-time feedback, most problems are often detected only in the later stages of experimentation, leading to delayed adjustments that compromise material performance stability.
Therefore, although extensive experimental and computational studies have successfully identified suitable process conditions to achieve target performance, a large number of promising combinations remain unexplored due to time and resource limitations. To overcome these constraints and to further explore the broad process parameter space, AL is utilized for process optimization tasks. AL addresses these challenges by dynamically collecting data and refining experimental strategies in real-time, efficiently exploring the complex and high-dimensional process parameter space, and making timely adjustments to maintain process stability. By continuously optimizing the workflow, AL mitigates the risk of delayed corrections and rapidly establishes precise correlations between material performance and process parameters, thereby enhancing both efficiency and reliability.

In catalyst development, AL excels at navigating the combinatorial explosion of multi-component formulations by efficiently locating high-performance regions within vast search spaces. For instance, the integration of evolutionary strategies into AL frameworks offers a systematic methodology~\cite{166}. By mimicking natural selection, these hybrid approaches mitigate the inefficiencies of conventional grid searches, identifying optimized structures within a limited number of experimental cycles. In the synthesis of higher alcohols from syngas, AL-integrated workflows have enabled the identification of high-performing catalyst-condition pairings from massive combinatorial spaces, while simultaneously elucidating the Pareto frontiers between key performance metrics~\cite{3}.
Target organic synthesis mirrors these advancements by focusing on the dual objectives of maximizing yield and integrating sustainability into chemical processes. In addressing challenging molecular transformations, AL simultaneously regulates multidimensional variables such as acid concentrations, thermal profiles, and solvent ratios, effectively resolving efficiency bottlenecks while upholding stringent green chemistry standards~\cite{37}.
When dealing with "black-box" reactions characterized by unclear mechanisms or extreme data scarcity, AL frameworks are uniquely capable of optimizing parameters to maximize the yield of desired products like light olefins from plastic waste pyrolysis~\cite{72}. To overcome the "cold-start" problem in reaction condition prediction, active transfer learning leverages knowledge from data-rich reaction types to predict conditions for related but data-scarce substrates, significantly improving predictive accuracy~\cite{119}. To transcend the limitations of reaction-specific models, universal AL strategies have been introduced, achieving efficient optimization without exhaustive screening by intelligently refining key variables~\cite{123,132}.

The application of AL in structural and functional materials facilitates the systematic optimization of processing parameters. For instance, physics-informed Bayesian optimization can systematically optimize processing parameters like heat-treatment temperature to maximize the phase transformation temperature in shape memory alloys~\cite{174}. Similarly, for perovskite nanocrystal synthesis, efficient goal-seeking algorithms leverage historical data to adaptively predict optimal reaction conditions~\cite{183}. AL is particularly effective in resolving multi-objective conflicts in advanced manufacturing. It demonstrates superior capability for the synergistic regulation of typically conflicting properties in additive manufacturing, discovering formulations with ideal mechanical trade-offs within very few iterations~\cite{171}. In processes like laser powder bed fusion, Bayesian optimization simultaneously maximizes strength and ductility, substantially reducing experimental workload~\cite{12}. Even in systems with high prediction uncertainty, AL can effectively identify experimental conditions that overcome conventional trade-off boundaries~\cite{51}.

In analytical chemistry and separation engineering, the implementation of AL is advancing experimental workflows toward dynamic optimization. In liquid chromatography method development, it enables the simultaneous tuning of parameters such as eluent ratio, flow rate, and gradient duration, enhancing separation performance while substantially reducing the time and expertise required for method establishment~\cite{175}. In membrane‑based separations, exemplified by vacuum membrane distillation for isopropanol recovery, optimization of critical parameters including feed temperature and transmembrane pressure has led to improved recovery efficiency~\cite{99}. In high‑precision industrial settings such as semiconductor manufacturing, AL facilitates human‑machine collaboration for optimizing intricate plasma‑etching processes: engineers drive strategic exploration, while algorithms efficiently perform parameter fine‑tuning. Through the co‑optimization of key variables—pressure, source power, and gas flow—this approach reduces both development cycles and associated costs~\cite{173}.

Although AL has achieved significant milestones in process optimization, its real-time responsiveness is often bottlenecked by the physical requirements of experimental workflows. The persistent reliance on manual intervention for sampling and analysis creates a feedback latency that prevents AL from operating at its full temporal potential. To mitigate this decoupling between digital reasoning and physical execution, there is a growing necessity for systems that can unify these two domains.

\subsection{Autonomous self-driving laboratories}
Traditional experimental workflows remain largely constrained by manual and fragmented processes~\cite{162}. In conventional practice, scientists often read experimental data by hand, record them in laboratory notebooks, manually transcribe them into spreadsheets, and then share the results with ML experts for further analysis. Such fragmented procedures severely limit research throughput, reproducibility, and scalability.
The emergence of SDLs represents a paradigm shift, transforming this fragmented approach into a cohesive, closed-loop ecosystem (Fig. \ref{fig:figApplication}f). By integrating automated high-throughput experimentation with intelligent AL algorithms, SDLs do not merely automate repetition; they enable the intelligent and systematic navigation of the expansive search space. Rather than aiming for full automation, modern SDLs emphasize a symbiotic relationship, creating rapid surrogate experiments and strategically placing human interventions to achieve an optimal balance between speed, flexibility, and interpretability.

In the transition of chemical synthesis toward automation and intelligence, SDLs have evolved beyond a mere laboratory auxiliary tool to become the central execution hub for constructing autonomous discovery loops. These platforms integrate automated experimental planning, synthesis, and characterization with real-time analysis and machine learning decision-making into a closed-loop ``experiment-learn-decide'' cycle, enabling autonomous exploration of vast chemical spaces. A pioneering case is an integrated robotic platform for chemical reaction discovery. This platform employs liquid-handling units to automatically prepare reagent combinations and carry out reactions, while online NMR and infrared spectroscopy provide real-time outcome analysis. ML models classify reactions as ``successful'' or ``unsuccessful'' based on the spectral data. The core of the platform lies in its closed-loop workflow: an initial set of random experiments builds a starting database; then, models such as linear discriminant analysis are used to predict the reactivity of unexplored combinations, and the experiments with the highest predicted reactivity are prioritized for the next round of validation. This active navigation strategy allows the system to achieve $>80\%$ prediction accuracy for over 1,000 reaction combinations after exploring only about 10\% of the space, leading to the discovery of several novel multicomponent reactions~\cite{181}. This work exemplifies the potential of such systems to dramatically accelerate the discovery of new chemical reactivity.
This paradigm extends further to more complex scenarios such as material synthesis and system optimization. In nanomaterial synthesis, microfluidic platforms have been used to extract interpretable physical insights into parameter-property relationships during the synthesis of silver nanoparticles~\cite{184,10}.
For processes with large search spaces, such as solubility screening and crystallization control, SDLs also demonstrate high exploration efficiency. For instance, in screening binary solvent systems for redox flow batteries, electrolyte formulations with solubilities exceeding 6.20 M were identified by sampling less than 10\% of the candidate space~\cite{15}. In crystallization studies, integrating robotic workflows with antisolvent vapor-assisted crystallization enabled phase diagram mapping and programmable control over crystal dimensionality with a very low sampling rate~\cite{76}. When exploring complex crystalline systems such as gigantic polyoxometalates, autonomous algorithms showed clear advantages over traditional human-experience-based methods, achieving higher crystallization success rates and exploring broader regions of phase space~\cite{182}.

The application of SDLs further extends to more challenging domains such as solid-state synthesis and thin-film fabrication. The A-Lab systematically analyzes failed experiments, attributing them to factors such as slow reaction kinetics, precursor volatility, amorphization, and computational inaccuracies, and extracts actionable suggestions for improvement. These analyses not only optimize subsequent experimental designs but also provide a direct basis for refining materials screening and synthesis strategies~\cite{31}. In thin-film material development, SDLs not only enable high-throughput, low-material-consumption screening of photostability across thousands of quaternary compositions~\cite{185}, but also navigate the Pareto front between conductivity and processing temperature through multi-objective optimization to address thermal compatibility issues in flexible electronics~\cite{179}.
The form of SDLs is evolving from fixed workstations toward flexible, anthropomorphic autonomous mobile robots. One such mobile robot for photocatalytic hydrogen production research is capable of navigating standard laboratory spaces and operating conventional instruments, thereby breaking the constraints of traditional automation islands~\cite{170}. Over eight days of continuous operation, the robot not only executed 688 experiments within a 10-dimensional variable space to achieve a sixfold increase in hydrogen production but also captured non-linear regulatory mechanisms of ionic strength and pH that are often overlooked by human intuition. This work signals the metamorphosis of SDLs from mere high-throughput tools into autonomous intelligent research entities capable of lab-scale perception and operation.
A recent breakthrough introduces the CRESt platform, a multimodal platform that distinguishes itself by integrating large multimodal models to synthesize diverse data streams such as literature text embeddings and microstructural images into the discovery loop. Unlike traditional unimodal systems, CRESt leverages Vision-Language Models to autonomously diagnose and correct experimental anomalies through real-time camera monitoring and natural language reasoning. This multimodal approach enabled the platform to explore an octonary chemical space and identify a state-of-the-art electrocatalyst with a 9.3-fold improvement in cost-specific performance~\cite{187}.

In summary, AL serves as the logical backbone of SDLs, transforming traditional linear workflows into dynamically evolving closed-loop systems by establishing an autonomous ``experiment-learn-decide'' cycle. Its core value lies in the use of surrogate models to evaluate uncertainty and potential gain within chemical spaces in real time, shifting conventional high-throughput experimentation from large-scale random screening toward goal-oriented heuristic exploration. Although current systems remain largely confined to parameter optimization for specific tasks, the gradual integration of agents equipped with reasoning capabilities promises to enhance the adaptability of future automated laboratories through the fusion of multimodal perception and physical prior knowledge. This evolution does not aim for absolute autonomy entirely devoid of human involvement; rather, it seeks to establish a more robust human-machine collaborative framework, enabling the system to handle routine exploratory tasks while assisting scientists in uncovering deeper underlying physicochemical mechanisms.

\begin{table*}[t!]
\caption{A summary of active learning applications in materials science, categorizing key domains by research focus, primary objectives, and relevant literature.}
\centering
\label{application_table}
\resizebox{1.0\linewidth}{!}{
\begin{tabular}{>{\centering\arraybackslash}m{3.5cm}
                  |>{\centering\arraybackslash}m{3.5cm}
                  |>{\centering\arraybackslash}m{6.0cm}
                  |>{\centering\arraybackslash}m{4.5cm}}

\toprule
Application & Focus & Goal & Reference \\
\hline

\multirow{4}[+23]{=}{\centering Accelerating Computational Simulations} & A General Framework & \multicolumn{1}{m{6.0cm}|}{\raggedright\arraybackslash Identifying stable structures and predicting their properties}& \cite{1,160,25,41,161} \\
\hhline{~|---}
& Phase Transition Simulation & \multicolumn{1}{m{6.0cm}|}{\raggedright\arraybackslash Capturing rare transition states and interfaces during phase transitions} &  \cite{33,30,110,135,142,127,128} \\
\hhline{~|---}
& Non-Equilibrium Transport Properties & \multicolumn{1}{m{6.0cm}|}{\raggedright\arraybackslash Bridging the timescale gap to enable accurate calculation of non-equilibrium transport properties} & \cite{130,159} \\
\hhline{~|---}
& Reactive MD & \multicolumn{1}{m{6.0cm}|}{\raggedright\arraybackslash Simulating transient reaction mechanisms and dynamic structural evolution in complex heterogeneous environments}& \cite{131,16,107,4,144,14,17,18,129} \\
\hline

\multirow{2}[+10]{=}{\centering Compositional Design Optimization}
& Alloy Design & \multicolumn{1}{m{6.0cm}|}{\raggedright\arraybackslash Regulating elemental stoichiometry and coordination environments to achieve targeted alloy design} & \cite{8,116,75,60,100,57,2,29,78,53,26,91,102,19} \\
\hhline{~|---}
& Molecular Design & \multicolumn{1}{m{6.0cm}|}{\raggedright\arraybackslash Achieving targeted molecular design through the precise substitution of functional groups and molecular backbones} & \cite{134,176,120,113} \\
\hline

\multirow{4}[+20]{=}{\centering Structural Optimization}
& Metallic \& Alloys & \multicolumn{1}{m{6.0cm}|}{\raggedright\arraybackslash  Refining atomic arrangements and surface configurations to optimize catalytic performance} & \cite{35,143} \\
\hhline{~|---}
& Inorganic \& Oxides & \multicolumn{1}{m{6.0cm}|}{\raggedright\arraybackslash  Modulating geometric structures and bonding environments to predict interfacial properties} & \cite{42,67} \\
\hhline{~|---}
& Porous Frameworks & \multicolumn{1}{m{6.0cm}|}{\raggedright\arraybackslash  Regulating topological features and pore architectures to enhance transport and stability} & \cite{122,50} \\
\hhline{~|---}
& Organic \& Hybrid Materials & \multicolumn{1}{m{6.0cm}|}{\raggedright\arraybackslash  Engineering molecular arrangements and microscopic textures to tune functional responses} & \cite{54,22,108,77,38} \\
\hline

\multirow{3}[+15]{=}{\centering Synthesis and Processing Optimization}
& Catalyst \& Organic & \multicolumn{1}{m{6.0cm}|}{\raggedright\arraybackslash Optimizing multi-variable conditions and formulations to maximize reaction yield and selectivity}& \cite{166, 3, 37, 72, 119, 123, 132} \\
\hhline{~|---}
& Functional Materials & \multicolumn{1}{m{6.0cm}|}{\raggedright\arraybackslash Balancing processing parameters to optimize functional and mechanical performance}& \cite{174,183,171,47,51,12} \\
\hhline{~|---}
& Chemical Analysis \& Separations & \multicolumn{1}{m{6.0cm}|}{\raggedright\arraybackslash Tuning operational variables to enhance separation efficiency and process reliability} & \cite{175, 99, 173} \\
\hline

Autonomous Self-Driving Laboratorie
& - & \multicolumn{1}{m{6.0cm}|}{\raggedright\arraybackslash Integrating robotic automation with AL to establish autonomous closed-loops for accelerated material discovery} & \cite{181,184,10,76,15,182,187} \\
\hhline{~|---}
\hline
\end{tabular}
}
\end{table*}

AL offers significant advantages across various domains of materials science. A summary of its key applications, categorized by different areas, objectives, and specific examples, is provided in Table~\ref{application_table}.

\section{Outlook}\label{challenge}
Despite the advances summarized above, the exploration of AL in materials science are still in their infancy. Several existing challenges must be addressed for the effective application of AL in this field to reach its full potential. Furthermore, there is also a lack of systematic benchmarking studies or practical tools to guide the implementation and evaluation of various AL strategies.
Tackling these challenges will require close collaboration between materials scientists and computer scientists across diverse research disciplines.

\subsection{Cold-start problem}
As discussed in Section~\ref{cold}, selecting the most promising initial data plays a crucial role in rapidly initiating experimental design optimization or computational simulation, which in turn has a strong impact on subsequent optimization directions and model performance. Existing strategies show great promise for addressing the cold-start problem and generally fall into three categories: random sampling, approaches that select initial samples based on domain knowledge or existing literature~\cite{54,58,100}, and those that identify representative samples through data distribution~\cite{72,114,116}. Despite these advances, the cold-start problem remains a considerable challenge in materials science.

In domain knowledge-guided methods, initial sample selection relies on expert experience and prior research insights, providing valuable guidance in the early stages of learning. However, these approaches are inherently limited by their dependence on subjective judgment, and discrepancies among experts can introduce bias and inconsistency. Using literature-derived data as initial samples can broaden the search space, but its effectiveness is often constrained by the availability and quality of the data. In practice, literature-based data may suffer from issues such as noise, missing entries, or inconsistencies compromising reliability and reproducibility. The challenge posed by incomplete and imperfect literature data can be partially mitigated through LLMs, which enable the automated retrieval and curation of information from the literature. By employing natural language processing and ML techniques, recent studies~\cite{polak2024extracting, peng2025ai, dagdelen2024structured} have demonstrated the extraction, filtering, and standardization of key information from large textual corpora, thereby systematically improving the quality and usability of literature-based initial datasets.

Another alternative approaches involve distribution-based AL strategies, which enhance model performance by selecting initial subset that effectively represent the overall materials space. This method addresses the inefficiencies of random sampling, particularly in large-scale datasets. However, its success depends heavily on the quality of data representation. Constructing physically meaningful chemical descriptors or extracting informative features is essential for capturing the intrinsic distribution of material space. Therefore, more efforts should be devoted to developing robust and physically relevant feature representations to further enhance the effectiveness of distribution-based AL in materials science.

Overall, although several attempts have been made to address the cold-start problem, the challenge has often been overlooked in materials science, with relatively few studies systematically tackling the issue. Existing efforts are often fragmented or application-specific, and lack unified theoretical frameworks and benchmark datasets to support broader adoption. Consequently, future research in materials science is expected to focus on developing more systematic and domain-specific strategies to overcome cold-start issues.

\subsection{Prior knowledge Integration}
Embedding the domain knowledge into the design of AL frameworks plays a central role in accelerating optimization efficiency. Fundamental chemical and physical principles such as symmetry, conservation laws, and thermodynamic stability define the boundaries of the admissible solution space. By integrating these priors into the AL framework as hard constraints, the search space can be systematically reduced and restricted to physically feasible regions, thereby improving data efficiency. Several studies \cite{28,104} have explored the domain knowledge at various stages of AL procedure, including feature construction, search space design and filtering, AL strategy design, and assisted AL.

\textbf{Feature construction.} Many existing studies rely on hand-crafted formulations that are closely coupled with intrinsic characteristics of the studied system and domain-specific expertise~\cite{54,53,29}.
Moreover, several studies utilize established chemical descriptors, such as SOAP and CBAD, to describe sample features, where the choice of descriptor critically affect data representation and subsequent sample selection~\cite{14,18,2}. Despite their successes, such hand-crafted features often lack flexibility, depend strongly on expert knowledge, and are typically tailored to specific material systems. Text-based~\cite{gomez2018automatic}
and graph-based representations~\cite{hu2019strategies} offer a more expressive means of capturing structural and relational information; however, they generally rely on well pre-trained models. Collectively, these limitations motivate the development of more flexible and generalizable representations that can better support subsequent learning and sampling in AL frameworks.

\textbf{Search space filtering.} Accurate domain knowledge should also be leveraged to refine the search space, thereby narrowing the optimization space in AL and accelerating the solution discovery process. For example, the search space can be divided into distinct regions based on physicochemical knowledge~\cite{28}, or parameter ranges can be constrained by incorporating prior knowledge~\cite{72,171}, such as imposing a carbon content limit of $\le 50\%$ to prevent the formation of brittle samples.

\textbf{AL strategies design.} Once the search space has been established, AL strategies are employed to identify the most informative samples for model improvement. Incorporating the domain knowledge into the design of AL strategies is another major challenge, thus ensuring that the selected samples comply with known feasibility constraints~\cite{142,14}.
For instance, hypothesis-driven frameworks have been developed that combine physical hypotheses with structured probabilistic models, and further incorporate reinforcement learning policies to dynamically guide experimental design and model validation~\cite{104}.
Similarly, the inclusion of physically meaningful terms derived from effective Hamiltonians within the sampling strategy serves to enforce feasibility during the selection process~\cite{142}.

\textbf{Assisted Active learning.} Beyond relying solely on AL strategies to select valuable samples, some studies directly integrate human expertise to jointly guide the selection process. This hybrid paradigm, commonly referred to as assisted AL~\cite{54,81,173,145}, combines expert knowledge with algorithmic strategies to collaboratively guide the data selection. In practice, experts typically intervene in the early stages of sample selection, after which AL algorithms perform fine-grained optimization. Such human-in-the-loop frameworks have demonstrated the potential to reduce experimental costs and mitigate the poor screening performance commonly observed during the early stages of model construction~\cite{81,173}. For instance, multi-stage expert collaboration schemes can refine molecular complexity models through iterative feedback. By prioritizing uncertain molecular pairs with comparable complexity for expert calibration, these methods emphasize subtle distinctions that significantly enhance labeling efficiency and model fidelity~\cite{145}.

Given the pivotal role of domain knowledge in the design and implementation of AL frameworks, further efforts are required
to tailor AL strategies to the requirements of specific scientific scenarios. However, such customization often comes at the cost of reduced generalizability, as strategies optimized for one system or task may not readily transfer to others. These challenges highlight a fundamental tension between domain specificity and methodological generality, underscoring the need for more adaptive and data-efficient AL frameworks that can balance physical consistency with broad applicability.

\subsection{Principled active learning configuration}
The performance of an AL approach hinges on several critical factors, including the choice of AL sampling strategy, the number of AL cycles, the budget allocated per cycle, the initial dataset size and others. Despite substantial progress in AL methodologies, selecting an appropriate algorithm and configuration for a novel system or application remains largely guided by educated guesses and subjective preferences. One major challenge is the absence of comprehensive guidelines or a unified framework to support informed and efficient decisions when configuring AL settings. Furthermore, the variability across various materials tasks impairs reproducibility, while the high computational or experimental cost renders exhaustive exploration of AL configurations impractical. These challenges collectively restrict the broader adoption of AL in materials science.

Insights from vision-based deep AL studies have noted that low-budget and high-budget regimes represent fundamentally different learning mechanisms, thus requiring opposite query strategies~\cite{hacohen2022active}. Specifically, uncertainty-based criteria are better suited to high-budget regimes, while representativeness-based selection is more effective under low-budget constraints. Although a systematic validation of this conclusion in materials science remains limited, emerging evidence suggests that the underlying principle is likely to generalize: representativeness plays a crucial role in low-budget settings, particularly in the early AL rounds~\cite{72,114,99}. This is often attributed to the limited capacity of task models to reliably estimate uncertainty at the initial stage, while training on more representative samples enables rapid improvements in model performance. By contrast, uncertainty-based AL approaches become more effective once decision boundaries are partially established, typically in high-budget regimes or later AL cycles~\cite{16,8}. Such uncertainty-driven methods are particularly valuable for probing unexplored regions of the input space and supporting extrapolative learning tasks.

In addition, foundational research into low-data discovery scenarios has demonstrated that the optimal acquisition function is critically dependent on both the underlying model and the dataset architecture~\cite{van2024traversing}. For DL-based approaches, exploitation-driven strategies and mutual information criteria typically identified the largest number of hits, although their relative effectiveness varied across datasets. By contrast, exploration-based strategies yielded the best performance for a state-of-the-art random forest baseline included for comparison. These findings indicate that the choice of AL selection function should be tailored to the specific model-dataset combination. Moreover, the number of AL iterations was found to have a substantial impact on performance: simple similarity-based selection methods often demonstrated competitive performance during the early learning stages. This suggests that a sufficient number of iterations may be required before surrogate models begin to exhibit clear advantages over simpler selection strategies.

Although no unified guidelines currently exist for selecting AL strategies in materials science, some studies have proposed practical solutions for choosing appropriate methods. In practice, some researchers employ small-scale datasets or systems as toy examples to compare different AL sampling strategies, thereby identifying suitable AL strategies for larger-scale tasks~\cite{15,18}. For instance, benchmark experiments conducted on a dataset of 98 known solvents demonstrated that Bayesian optimization serves as a robust and efficient method for accelerating the identification of candidates with specific solubility requirements~\cite{15}. Similarly, evaluations performed on small water systems indicate that configurations chosen through similarity or distance based selectors exhibit a more uniform distribution. This provides a scalable pathway for selecting effective AL protocols in complex materials discovery workflows~\cite{18}.

In future, there is a pressing need for systematic benchmarks that rigorously compare AL strategies across different experimental settings and materials-specific tasks, while also developing unified, framework-level guidelines for their practical deployment in materials science. Such efforts would enable a comprehensive evaluation of the strengths, limitations and applicability of existing AL approaches across varying data regimes, model assumptions and task objectives, while providing principled guidance tailored to specific materials discovery and design problems. More broadly, they would help establish realistic expectations for performance gains and accelerate the adoption of AL as a scalable and reliable paradigm for data-driven materials science.

\subsection{Materials-oriented AL tools}
Although a growing body of work has successfully demonstrated the application of AL to individual materials tasks, these successes have largely remained domain-specific case studies. A key challenge is to consolidate such isolated successes into autonomous tools that systematically accelerate research efficiency, foster interdisciplinary collaboration, and enable efficient exploration of vast chemical spaces for materials discovery.

Several AL tools have been developed within the computational science community, including ALiPy~\cite{TLHalipy}, modAL~\cite{modAL2018}, and Trieste~\cite{trieste2023}. However, these tools typically require substantial adaptation to be applicable to materials-specific scenarios. Thus, it is essential to develop AL tools tailored for materials research, particularly for domain scientists without extensive AI expertise. Bgolearn \cite{cao2026bgolearnunifiedbayesianoptimization} has been developed as a Python framework that makes BO accessible and practical for materials research. Nevertheless, further strategies should be integrated into AL tools, not only to offer intuitive, user-friendly interfaces but also to provide comprehensive guidance that facilitates the materials-oriented workflows. We envision that this interdisciplinary cross-pollination, drawing on advances in computational science and materials science, will foster innovative discoveries, broaden the scope of scientific understanding, and potentially transform paradigms in scientific research.

\subsection{Large language models and autonomous agents}

Materials science is undergoing a profound transition from empirically driven trial-and-error approaches toward data-driven and intelligent discovery paradigms. However, current AI applications remain highly fragmented. Tasks such as data extraction, property prediction, and experimental decision-making are typically executed by disparate models, hindering the realization of an end-to-end, closed-loop workflow spanning conceptual design, candidate screening, and experimental validation. In this context, LLMs, such as GPT \cite{openai2024gpt4technicalreport} and BERT \cite{devlin2019bert}, have demonstrated powerful capabilities in semantic understanding and knowledge integration. By unifying heterogeneous information representations and reasoning processes, general-purpose LLMs offer a promising pathway toward the development of general materials intelligence. In addition, the emergence of autonomous agents further advances this vision by enabling goal-driven decision-making, planning, and execution through the coordinated invocation of domain-specific models, multimodal databases, and experimental robotics, thereby moving closer to truly autonomous materials discovery.

Nevertheless, in contrast to high-resource modalities such as natural language and vision, materials science continues to face fundamental challenges arising from the scarcity of high-quality labeled data, limited reproducible experimental feedback, and multi-scale heterogeneity. Leveraging data-efficient paradigms, particularly AL, to guide LLMs and autonomous agents toward high-value sampling in critical regions of the design space remains a central challenge for building reliable and scalable materials intelligence systems. The realization of such systems depends on bridging the gap between physics-grounded reasoning, knowledge-informed sampling, and the efficient orchestration of automated workflows.

\backmatter

\bmhead{Acknowledgments}

This work is supported by National Science and Technology Major Project (2023ZD0121101), National University of Defense Technology (ZZCX-ZZGC-01-04) and Major Fundamental Research Project of Hunan Province (2025JC0005).

\begin{appendices}

\end{appendices}

\bibliography{sn-bibliography}

\end{document}